%% file: main.tex
\documentclass[manuscript,nonacm,screen]{acmart}
\AtBeginDocument{%
  }

\setcopyright{acmlicensed}
\copyrightyear{2027}
\acmYear{2027}
\acmDOI{XXXXXXX.XXXXXXX}
\acmConference[Conference acronym 'XX]{}{June 03--05,
  2027}{Woodstock, NY}
\acmISBN{978-1-4503-XXXX-X/2027/06}

\usepackage{float}
\usepackage{array}

\usepackage{xcolor}    
\usepackage{colortbl}  
\usepackage{listings}  
\usepackage{graphicx}
\usepackage{booktabs}
\usepackage[table]{xcolor}

\definecolor{withdrawalbluebg}{RGB}{229,238,250}
\definecolor{confrontredbg}{RGB}{250,230,230}
\definecolor{withdrawalblue}{RGB}{30,60,130}
\definecolor{confrontred}{RGB}{150,30,30}

\definecolor{promptgray}{RGB}{246,246,241}

\begin{document}




\title[Can LLMs identify and repair ruptures? Comparison between clinician practices and LLM behaviors]{Can LLMs identify and repair ruptures? Comparison between clinician practices and LLM behaviors}


\author{Jeongah Lee}
\orcid{0000-0002-5714-5521}
\affiliation{%
  \institution{University of Massachusetts Amherst}
  \city{Amherst}
 \state{MA}
  \country{USA}
}
\email{jeongahlee@umass.edu}

\author{Joy Qiuyue Zhong}
\orcid{0009-0006-4328-2329}
\affiliation{%
  \institution{University of Massachusetts Amherst}
  \city{Amherst}
 \state{MA}
  \country{USA}
}
\email{qzhong@umass.edu}

\author{Drishti Goel}
\orcid{0009-0000-6713-9240}
\affiliation{%
 \institution{University of Illinois Urbana-Champaign}
 \city{Urbana}
 \state{IL}
 \country{USA}}
 \email{drishti4@illinois.edu}

\author{Violeta J. Rodriguez}
\orcid{0000-0001-8543-2061}
\affiliation{%
 \institution{University of Illinois Urbana-Champaign}
 \city{Champaign}
 \state{IL}
 \country{USA}}
 \email{vjrodrig@illinois.edu}

\author{Dong Whi Yoo}
\orcid{0000-0003-2738-1096}
\affiliation{%
 \institution{Indiana University Indianapolis}
 \city{Indianapolis}
 \state{IN}
 \country{USA}}
 \email{dy22@iu.edu}

\author{Koustuv Saha}
\orcid{0000-0002-8872-2934}
\affiliation{%
 \institution{University of Illinois Urbana-Champaign}
 \city{Urbana}
 \state{IL}
 \country{USA}}
 \email{ksaha2@illinois.edu}

\author{Ravi Karkar}
\orcid{0000-0003-1467-4439}
\affiliation{%
 \institution{University of Massachusetts Amherst}
 \city{Amherst}
 \state{MA}
 \country{USA}}
 \email{rkarkar@umass.edu}

\renewcommand{\shortauthors}{Lee et al.}

\newcommand{\jl}[1]{\textcolor{blue}{\textbf{[Jasmine: #1]}}}
\newcommand{\jz}[1]{\textcolor{magenta}{\textbf{[Joy: #1]}}}
\newcommand{\dg}[1]{\textcolor{cyan}{\textbf{[Drishti: #1]}}}
\newcommand{\profv}[1]{\textcolor{red}{\textbf{[Prof.V: #1]}}}
\newcommand{\ravi}[1]{\textcolor{orange}{\textbf{[Ravi: #1]}}}
\newcommand{\tiered}{\textsc{tiered selection process}}

\newcommand{\participantquote}[2]{{\small \textit{#1}~#2}}
\newcommand{\Aone}{\textsc{A1}}
\newcommand{\Atwo}{\textsc{A2}}
\newcommand{\Athree}{\textsc{A3}}

\newcommand{\Ione}{\textsc{I1}}
\newcommand{\Itwo}{\textsc{I2}}

\newcommand{\Cone}{\textsc{C1}}
\newcommand{\Ctwo}{\textsc{C2}}
\newcommand{\Cthree}{\textsc{C3}}

\newcommand{\Done}{\textsc{D1}}
\newcommand{\Dtwo}{\textsc{D2}}
\newcommand{\Dthree}{\textsc{D3}}
\newcommand{\Dfour}{\textsc{D4}}

\newcommand{\DEone}{\textsc{E1}}
\newcommand{\Sone}{\textsc{N1}}

\newcommand{\experts}{experts}
\newcommand{\Experts}{Experts}
\newcommand{\expert}{expert}
\newcommand{\Expert}{Expert}

\newcommand{\repairing}{resolving}
\newcommand{\repair}{resolution}
\newcommand{\repairs}{resolutions}
\newcommand{\Repair}{Resolution}
\newcommand{\Repairs}{Resolutions}
\newcommand{\repairResolve}{resolve}

\newcommand{\resolution}{resolution}
\newcommand{\Resolution}{Resolution}

\newcommand{\pAone}{(\Aone)}
\newcommand{\pAtwo}{(\Atwo)}
\newcommand{\pAthree}{(\Athree)}

\newcommand{\pIone}{(\Ione)}
\newcommand{\pItwo}{(\Itwo)}

\newcommand{\pCone}{(\Cone)}
\newcommand{\pCtwo}{(\Ctwo)}
\newcommand{\pCthree}{(\Cthree)}

\newcommand{\pDone}{(\Done)}
\newcommand{\pDtwo}{(\Dtwo)}
\newcommand{\pDthree}{(\Dthree)}
\newcommand{\pDfour}{(\Dfour)}

\newcommand{\pDEone}{(\DEone)}
\newcommand{\pSone}{(\Sone)}


\definecolor{tableheader}{rgb}{0.404, 0.663, 0.812}   
\definecolor{tablehighlight}{rgb}{0.800, 0.800, 0.800} 

\newcommand{\headerrow}{\rowcolor{tableheader}}
\newcommand{\highlightrow}{\rowcolor{tablehighlight}}

\newenvironment{styledtable}[1]{%
  \setlength{\tabcolsep}{4pt}%
  \renewcommand{\arraystretch}{1.25}%
  \arrayrulecolor{tableheader}%
  \begin{tabular}{#1}%
}{%
  \end{tabular}%
  \arrayrulecolor{black}%
}

\newcommand{\headcol}{\rowcolor{tableheadcolor}}
\newcommand{\rowcol}{\rowcolor{tablerowcolor}}
\newcommand{\rowcolmedium}{\rowcolor{tablerowcolor2}}
\newcommand{\rowcollight}{\rowcolor{tabgray}}
\newcommand{\rowcolgroup}{\rowcolor{groupgray}}

\newcommand{\tablestyle}{\sffamily\footnotesize}
\newcommand{\tabletitlestyle}{\sffamily}

\newcommand{\tablegroup}[1]{%
  \rowcolor{groupgray}%
  \multicolumn{5}{l}{\textbf{#1}}%
}

\begin{abstract}

Ruptures represent common albeit critical moments in interaction where relational alignment breaks down, making them essential for evaluating AI where trust and engagement matter most.
In a scenario-driven empirical study, we examined the performance of three LLMs at identifying and resolving ruptures across 21 mental health conversations and 22 experts' evaluation of the strategies.
For identification, LLMs relied on explicit linguistic cues within single turns whereas experts integrated implicit, relational, and contextual information across the conversation. 
For resolution, LLMs tended to produce more directive and scripted responses whereas experts adopted process-oriented strategies such as validation, open-ended exploration, and psychoeducation.
Overall, LLMs showed higher agreement with predefined labels in identification, but not in resolution where experts rated their responses only moderately effective, with consistent limitations in timing, depth, and contextual sensitivity. 
We discuss implications for the design of mental health conversational agents emphasizing relational awareness, pacing, and human-in-the-loop support.


\end{abstract}

\ccsdesc[500]{Human-centered computing~Empirical studies in HCI}
\ccsdesc[300]{Applied computing~Health informatics}


\keywords{mental health, mental health experts, mental health chatbots, conversational AI, large language models, rupture, human-AI interaction}


\maketitle

\input{sections/1_introduction.tex}
\input{sections/2_related_work.tex}

\input{sections/3_design}

\input{sections/4_methods}
\input{sections/5_1_findings_identification}

\input{sections/5_2_findings_resolution}

\input{sections/5_3_findings_evaluation}

\input{sections/5_4_findings_survey}
\input{sections/6_Discussion}

\input{sections/7_conclusion}

\begin{acks}
This work was supported in part by the National Institute on Aging of the National Institutes of Health under Award Number P30AG073105 and the Jump ARCHES endowment through the Health Care Engineering Systems Center at the University of Illinois and the OSF Foundation.
\end{acks}

\bibliographystyle{ACM-Reference-Format}
\bibliography{main}

\clearpage
\appendix
\input{sections/appendix}

\end{document}

%% file: sections/1_introduction.tex
\section{Introduction}
Large language model-enabled AI conversational systems (chatbots) are increasingly used in emotionally sensitive and high-stakes contexts, including mental health support~\cite{yoo2026ai, casu2024ai, wester2024chatbot, lee2023exploring}, caregiver guidance~\cite{shi2025mapping, goel2026rubrix, shi2025balancing}, patient-facing healthcare~\cite{cuadra2024digital, yang2024talk2care}, and personal reflection, such as journaling and self-tracking~\cite{kim2024diarymate, kim2024mindfuldiary, chopra2025engagements}.
In these settings, users turn to AI systems for more than information, including emotional support, guidance, and reassurance~\cite{inkster2018empathy, fitzpatrick2017delivering, skjuve2022longitudinal, xu2025digital}. As a result, interaction effectiveness depends on the quality of the perceived relationship between users and AI systems. Prior work identifies this relationship as central to user engagement and trust~\cite{beatty2022evaluating, xu2025digital, darcy2021evidence, skjuve2022longitudinal}.

Development has moved well beyond clinical research teams. General-purpose assistants, lightweight application wrappers, and configurable personas have lowered the barriers to deploying a system that functions as mental health support, often without clinical training, institutional oversight, or an evaluation plan ~\cite{casu2024ai, haran2026checklist, yoo2026ai}.
Relational breakdown, however, is often overlooked by those building these systems. Existing safety mechanisms are largely oriented toward explicit and acute risk, such as crisis language or overtly harmful content~\cite{haran2026checklist, li2023systematic}. Yet a response can be fluent, policy-compliant, and superficially supportive, while still leaving the user feeling dismissed, misread, or unsupported.
Psychotherapy research provides a well-established framework for understanding these moments. Breakdowns are conceptualized as \textit{ruptures}, defined as critical moments when relational alignment between the user and the system becomes strained or disrupted~\cite{eubanks2022rupture, safran1996resolution, eubanks2018alliance}. 
Failures to identify and resolve ruptures can intensify user distress and deepen disengagement~\cite{yang2020re, gulay2026relational}, with consequences ranging from platform abandonment~\cite{o2022rethinking, al2025navigating} to more severe outcomes, including emotional dependency, social isolation, and self-harm~\cite{ortutay2025openai, brittain2026character, patel2024harmful}.

Despite these risks, existing evaluation frameworks for AI systems focus mainly on broad aspects such as empathy and perceived alliance~\cite{beatty2022evaluating, xu2025digital}, safety~\cite{haran2026checklist}, and harmful or inappropriate responses~\cite{li2023systematic, casu2024ai}. 
As a first step toward studying rupture identification in multi-turn conversations, we evaluate LLMs using fixed, expert-designed scenarios that contain clinically grounded rupture types. This scenario-based approach allows us to compare LLM responses in a controlled setting without claiming to evaluate live conversation monitoring.
We still have limited understanding of LLMs’ capabilities and limitations in identifying and resolving ruptures, particularly when evaluated in relation to how mental health \experts{} interpret and respond to the same interactions.
To address this gap, our work seeks to address the following research questions (RQs):

\begin{enumerate}
    \item \textbf{RQ1.} How do LLMs compare to mental health \experts{} in identifying rupture types, and what interpretive patterns characterize \experts{}' assessments of LLM identification?

    \item \textbf{RQ2.} How do mental health \experts{} resolve ruptures, and how do they evaluate LLM-generated \repair{} strategies?

\end{enumerate}

To address these questions, we evaluate rupture handling through 21 scenarios grounded in the Rupture Resolution Rating System (3RS)~\cite{eubanks2022rupture} --- a theory-driven, empirically validated, and behaviorally operationalized framework for identifying alliance ruptures and repair processes in psychotherapy. \Experts{} were asked to identify rupture types in seven scenarios, provide their own \repair{} strategies, and evaluate the effectiveness of the \repair{} strategies generated by three LLMs (ChatGPT 5.4 Thinking, Claude Sonnet 4.6 Extended Thinking, and Gemini 3 Thinking).
Our findings showed that high identification agreement co-occurred with only moderate ratings of resolution effectiveness.
Although LLMs achieved higher agreement with predefined reference labels in identifying rupture types in structured scenarios, this strength did not extend to resolution. \Experts{} found LLM-generated resolution strategies only moderately effective, with recurring concerns about timing, depth, and contextual fit.
Qualitative analysis revealed systematic differences in how rupture identification and \repair{} were carried out by LLMs and \experts. LLM outputs were largely driven by explicit linguistic cues within individual turns, whereas \experts{} drew on implicit, relational, and contextual signals across the broader interaction, treating ambiguity as an inherent part of the interpretive process. For rupture \repair{}, \experts{} adopted adaptive, process-oriented strategies such as validation, open-ended exploration, and collaboration, while LLMs tended toward narrower, more scripted responses that moved prematurely toward problem-solving.
Taken together, these results show that a system can appear safe and competent by current evaluation measures while still failing at the relational work that makes support helpful --- a gap that responsible deployment in this domain must account for.
Overall, this paper makes the following contributions:

\begin{enumerate}
    \item \textbf{An empirical framework for evaluating rupture identification and \repair{} in mental health conversations.}
    We provide a structured, scenario-driven methodology grounded in psychotherapy theory for systematically examining rupture handling. This offers developers and evaluators, including those without clinical training, a concrete instrument for examining relational failure before a system reaches users.

    \item \textbf{A characterization of the reasoning gap between LLMs and \experts{} in rupture identification and \repair{}.} We identify systematic gaps in how ruptures are interpreted and addressed. \Experts{} use implicit, relational, and contextual cues, while LLMs rely mainly on surface-level linguistic cues.

    \item \textbf{Design implications for conversational AI to support mental health practice.}
    We outline potential roles for conversational AI in mental health settings, emphasizing support for context-sensitive, adaptive \repair{} processes, and clearly bounded human-in-the-loop use in high-stakes interactions. We frame these as responsible design requirements rather than optional features, including explicit role boundaries, communicated uncertainty, and defined escalation pathways to human support.
\end{enumerate}

\noindent Rather than advocating for the replacement of human therapists, this work aims to characterize the capabilities and failure modes of LLMs in mental health contexts, motivated by their growing real-world use and the need to better understand their risks and limitations.

%% file: sections/2_related_work.tex
\section{Related Work}
We situate this work in research on conversational AI for mental health support (Section~\ref{sec:hai_mh_rw}) and on conversational breakdown and \repair{} in human-AI interaction (Section~\ref{sec:hai_rw}). We then provide the psychological background on the concept of rupture (Section~\ref{sec:rupture_rw}).

\subsection{Conversational AI for Mental Health Support}
\label{sec:hai_mh_rw}

Conversational agents have been used for mental health support since well before the current generation of language models. Early systems were rule-based and clinically scripted: Woebot delivered structured Cognitive Behavioral Therapy (CBT) exercises through decision-tree dialogue~\cite{darcy2021evidence}, and Wysa combined scripted CBT content with free-text input to support self-guided reflection~\cite{beatty2022evaluating}. Such systems were mostly developed by clinical research teams, evaluated against symptom measures, and deliberately constrained in what they could say~\cite{casu2024ai, li2023systematic}. Because the system could only select from a fixed set of pre-authored responses, any mismatch between user input and the script surfaced visibly---as an irrelevant reply or a stalled dialogue---and responsibility for what was said rested with the clinicians who authored the script, not with the system generating it in the moment.

LLMs remove that script: responses are generated live rather than selected from pre-approved content, so a mismatch no longer breaks the dialogue but instead produces a fluent, plausible-sounding reply, and no one has reviewed that specific response in advance~\cite{casu2024ai, wester2024chatbot}. Development has correspondingly moved beyond clinical research teams: general-purpose assistants, lightweight application wrappers, and configurable personas make it easy to deploy a system that functions as mental health support without clinical training, institutional oversight, or an evaluation plan~\cite{haran2026checklist, yoo2026ai}, and users increasingly approach such systems directly rather than through a clinically sanctioned pathway~\cite{lee2023exploring, chopra2025engagements}.

Conversational AI now occupies a relationally significant role across a widening range of settings: LLM-supported journaling and self-reflection, where systems scaffold users in articulating and revisiting emotional experience~\cite{kim2024mindfuldiary, kim2024diarymate, chopra2025engagements,saha2026ai}; caregiver guidance, where agents mediate emotionally demanding care work~\cite{shi2025mapping, shi2025balancing, goel2026rubrix}; patient-facing communication, where systems such as Talk2Care elicit and relay health information over ongoing interactions~\cite{yang2024talk2care, cuadra2024digital}; and companion applications, where users sustain emotionally consequential relationships over months of use~\cite{skjuve2022longitudinal, al2025navigating}. Across these settings, users turn to the system for reassurance, validation, and guidance, so effectiveness depends on the quality of the perceived relationship rather than on the accuracy of any single response~\cite{beatty2022evaluating, xu2025digital, darcy2021evidence}.

However, evaluation practice has not tracked this shift. Existing frameworks rely on broad relational constructs such as empathy and perceived alliance~\cite{beatty2022evaluating, xu2025digital}, safety checklists~\cite{haran2026checklist}, and detection of harmful content~\cite{li2023systematic, casu2024ai}. These measures are applied at the level of a single response or a fixed symptom outcome, rather than tracking how the quality of the interaction develops or degrades over the course of a conversation~\cite{cuadra2024illusion}. They also rarely benchmark system behavior against the judgment of mental health \experts, so it remains unclear whether what a system treats as an acceptable response would be recognized as adequate by someone trained to notice when a client is disengaging or losing trust in the process. Our study addresses this gap by directly comparing how LLMs and mental health experts identify and respond to such moments of relational difficulty over the course of a conversation.

\subsection{Conversational Breakdowns and \Repair{} in Human-AI Interaction}
\label{sec:hai_rw}

Prior human-AI interaction research has investigated conversational breakdown and \repair{}, developing foundational frameworks and vocabularies to analyze interactional failures~\cite{luger2016like}. 
However, this body of work has mostly operated under a task-oriented paradigm, spanning contexts such as voice assistants for information retrieval~\cite{luger2016like, porcheron2018voice}, in-car navigation~\cite{meck2024failing}, recommendation~\cite{bernard2024identifying}, and group facilitation~\cite{do2023err}. 
Within these settings, breakdown is conceptualized as a disruption to task progress where a system misunderstands intent, provides unhelpful information, or stalls workflow execution.
Conversation analytic studies demonstrate that when such disruptions occur, users bear the burden of progressivity through repeated reformulation~\cite{porcheron2018voice, fischer2019progressivity}, progressively adapting their linguistic choices to match system capabilities~\cite{mavrina2022alexa, dippold2023can}.

Complementing user-side adaptation, system-side HCI research has treated \repair{} primarily as a design problem focused on grounding and clarification~\cite{ashktorab2019resilient}. 
Subsequent work expanded these strategies to include apology, explanation, fallback responses, and redirection~\cite{alghamdi2024system, bernard2024identifying}, organizing them into broader taxonomies of confirmation, information provision, and social \repair{}~\cite{benner2021you, higashinaka2021overview}. 
These techniques range from prompting user rephrasing to explicit failure acknowledgment~\cite{ashktorab2025s}, with recent studies extending them across diverse modalities~\cite{li2020multi, do2023err, tommy2026conversational}. 
Yet these \repair{} mechanisms are designed to resume task progress, not to prevent the frustration, disengagement, and trust erosion that follow when a relational---rather than functional---expectation goes unmet~\cite{o2022rethinking, kahr2024trust}. 

This conventional framework, however, breaks down in mental health and supportive dialogue, where the core objective is not informational exchange or action execution, but the therapeutic alliance itself. 
In these contexts, a conversational failure is often a relational rupture---a perceived breakdown in attunement, validation, or emotional collaboration~\cite{safran2000resolving}---which typically exhibits none of the surface-level errors that existing taxonomies detect. 
A conversational agent can output an utterance that is grammatically sound, factually accurate, and contextually coherent, yet still leave a user feeling invalidated or alienated. 
Because conventional breakdown metrics look for stalled tasks and conventional \repair{} mechanisms aim to resume them, neither is equipped to diagnose or resolve relational dissonance. 
Our work bridges this gap by shifting the analytical lens from well-formedness and task progress to relational \repair{}, investigating how conversational agents identify and mend the nuanced failures that functional frameworks overlook.

\subsection{Therapeutic Alliance, Rupture, and \Repair{}}
\label{sec:rupture_rw}

Psychotherapy research provides a well-developed framework for understanding relational breakdowns in mental health conversations. At the center of this framework is the \textit{therapeutic alliance}, defined as agreement on goals, agreement on tasks, and an emotional bond between therapist and client~\cite{bordin1979generalizability}. The alliance is consistently linked to better outcomes across therapy modalities~\cite{karver2018meta}. Because therapy depends on collaboration and trust, disruptions in the alliance are central to understanding when and why therapeutic interactions become strained.

These disruptions are conceptualized as \textit{ruptures}. Ruptures are moments when the therapeutic relationship becomes strained or misaligned, often expressed through withdrawal, avoidance, compliance without agreement, dissatisfaction, or resistance~\cite{safran1996resolution, safran2000resolving}. Ruptures are considered inevitable components of psychotherapy and can provide opportunities for therapeutic change when effectively addressed~\cite{safran2000resolving}. Meta-analytic evidence also shows that successful rupture \repair{} is associated with improved treatment outcomes and reduced dropout~\cite{eubanks2018alliance}. Ruptures are therefore clinically meaningful moments that require careful recognition and response.

The Rupture Resolution Rating System (3RS) provides a structured approach to identifying rupture markers (Table~\ref{tab:rupture-tiered-abbrev})~\cite{eubanks2022rupture}. However, identifying ruptures alone does not explain how they are resolved. In practice, \repair{} is a context-dependent, multi-turn process that restores relational alignment through ongoing interpretation of interactional and relational cues~\cite{safran2011repairing, eubanks2022rupturerepair}.
Although the 3RS provides a structured framework for identifying and resolving rupture markers, it was developed for human therapists and human therapeutic relationships. As conversational agents are increasingly used in roles that involve emotional support and ongoing interaction~\cite{darcy2021evidence, skjuve2022longitudinal, beatty2022evaluating, lee2020designing}, it remains unclear how this psychotherapy-based framework applies to AI-mediated mental health conversations. In particular, little is known about whether LLMs identify rupture markers in ways that align with clinically grounded labels, or how their resolution strategies differ from those of mental health \experts. To address this gap, we build on the 3RS and use theory-driven scenarios to examine how LLM-based systems identify and respond to ruptures in mental health conversations.

Taken together, mental health AI has grown relationally significant (Section~\ref{sec:hai_mh_rw}), yet neither HCI's breakdown frameworks (Section~\ref{sec:hai_rw}) nor psychotherapy's rupture theory (Section~\ref{sec:rupture_rw}) has been used to examine how LLMs handle relational strain against expert judgment. We address this gap by comparing how LLMs and mental health \experts{} identify and resolve ruptures in the same conversations.

\begin{table*}[tbh]
\centering
\footnotesize
\sffamily
\setlength{\tabcolsep}{4pt}
\caption{Rupture taxonomy used in the study. The six high-level categories are grouped into two rupture classes, withdrawal and confrontation, with subtype abbreviation codes used in later analyses.}
\begin{tabular}{p{1.8cm} p{1.8cm} p{1.8cm} p{1.8cm} p{2.9cm} p{3.0cm}}
\toprule
\multicolumn{4}{c}{\textbf{Withdrawal}} 
& \multicolumn{2}{c}{\textbf{Confrontation}} \\
\cmidrule(lr){1-4}\cmidrule(lr){5-6}
{\raggedright \textbf{Avoidance}\par} 
& {\raggedright \textbf{Incongruence}\par} 
& {\raggedright \textbf{Disengagement}\par} 
& {\raggedright \textbf{Self-Directed Negativity}\par} 
& {\raggedright \textbf{Control}\par} 
& {\raggedright \textbf{Dissatisfaction}\par} \\
\midrule

{\raggedright Abstract communication \pAone\par}
& {\raggedright Deferential \pIone\par}
& {\raggedright Minimal response \pDEone\par}
& {\raggedright Self-criticism \pSone\par}
& {\raggedright Rejecting intervention \pCone\par}
& {\raggedright Complaint about progress \pDone\par}
\\

{\raggedright Avoidant storytelling \pAtwo\par}
& {\raggedright Content--affect split \pItwo\par}
&
&
& {\raggedright Defensiveness \pCtwo\par}
& {\raggedright Complaint about chatbot \pDtwo\par}
\\

{\raggedright Denial \pAthree\par}
&
&
&
& {\raggedright Efforts to control \pCthree\par}
& {\raggedright Complaint about therapy activities \pDthree\par}
\\

&
&
&
&
& {\raggedright Complaint about therapy parameters \pDfour\par}
\\

\bottomrule
\end{tabular}
\label{tab:rupture-tiered-abbrev}
\end{table*}

%% file: sections/3_design.tex
\section{Design}

We designed a study to examine how LLMs identify and \repairResolve{} ruptures in mental health conversation scenarios. 
We first describe how rupture scenarios were constructed (Section~\ref{sec:scenario-construction}), then how \repair{} options were generated (Section~\ref{sec:repair-options-generation}), and finally how both were organized into storyboards for the study (Section~\ref{sec:storyboard-consturction}).

\subsection{Rupture Scenario Construction}
\label{sec:scenario-construction}
We constructed 21 rupture scenarios based on the Rupture Resolution Rating System (3RS) manual~\cite{eubanks2022rupture}. We selected the 3RS because it provides a theory-driven, empirically validated, and behaviorally defined framework for identifying and resolving alliance ruptures. As a widely used observational coding system in psychotherapy, the 3RS offers fine-grained behavioral markers and structured \repair{} strategies grounded in clinical theory~\cite{eubanks2015rupture}. This allowed us to anchor our evaluation in a clinically grounded, process-oriented model.

Each of the 21 consists of approximately five-turn chatbot dialogues that contain one or more rupture markers. An example scenario is shown in Figure~\ref{fig:storyboards} (A) and full scenarios are provided in supplementary material.
The 3RS defines 14 rupture markers across withdrawal and confrontation classes. In order to reduce participants' cognitive load we reorganized these markers into six high-level categories (Table~\ref{tab:rupture-tiered-abbrev}). We then constructed 14 single-rupture scenarios to cover all markers and 7 mixed-rupture scenarios to capture more complex interaction patterns. 
All scenarios were adapted from the 3RS and translated into human--chatbot dialogue, where the therapist role is represented by the chatbot. Mixed rupture scenarios were created by combining commonly co-occurring rupture types, where a primary rupture drives the interactional strain and a secondary rupture modifies or amplifies it. Reference rupture labels are shown in Table~\ref{tab:scenario_gt}.
All scenarios were refined for clinical grounding and conversational plausibility and were validated by a clinician on the team.

\begin{table*}[t]
    \centering
    
    {\sffamily \footnotesize
    \setlength{\tabcolsep}{4pt}

    \caption{Reference rupture subtypes across the three storyboards. Each storyboard contains seven scenarios designed to cover diverse rupture types.}
    \label{tab:scenario_gt}
    \begin{tabular}{
        p{0.035\textwidth} p{0.255\textwidth}
        p{0.035\textwidth} p{0.255\textwidth}
        p{0.035\textwidth} p{0.255\textwidth}}
        \toprule
        \multicolumn{2}{c}{\textbf{Storyboard A}} 
        & \multicolumn{2}{c}{\textbf{Storyboard B}} 
        & \multicolumn{2}{c}{\textbf{Storyboard C}} \\
        \cmidrule(lr){1-2}\cmidrule(lr){3-4}\cmidrule(lr){5-6}
        \textbf{ID} & \textbf{Rupture Type} 
        & \textbf{ID} & \textbf{Rupture Type} 
        & \textbf{ID} & \textbf{Rupture Type} \\
        \midrule

        S1 & {\raggedright Denial\par}
        & S8  & {\raggedright Minimal response\par}
        & S15 & {\raggedright Complaint about therapy parameters\par} \\
        
        S2 & {\raggedright Rejecting intervention\par}
        & S9  & {\raggedright Complaint about chatbot\par}
        & S16 & {\raggedright Efforts to control\par} \\
        
        S3 & {\raggedright Abstract communication\par}
        & S10 & {\raggedright Avoidant storytelling\par}
        & S17 & {\raggedright Avoidant storytelling and complaint about chatbot\par} \\
        
        S4 & {\raggedright Self-criticism and rejecting intervention\par}
        & S11 & {\raggedright Complaint about progress\par}
        & S18 & {\raggedright Content--affect split\par} \\
        
        S5 & {\raggedright Deferential\par}
        & S12 & {\raggedright Defensiveness\par}
        & S19 & {\raggedright Complaint about chatbot and content--affect split\par} \\
        
        S6 & {\raggedright Deferential and complaint about progress\par}
        & S13 & {\raggedright Minimal response and defensiveness\par}
        & S20 & {\raggedright Complaint about therapy activities\par} \\
        
        S7 & {\raggedright Self-criticism\par}
        & S14 & {\raggedright Complaint about therapy activities and abstract communication\par}
        & S21 & {\raggedright Efforts to control and complaint about therapy parameters\par} \\
        \bottomrule
    \end{tabular}
    }
    
\end{table*}

\subsection{\Repair{} Options Generation}
\label{sec:repair-options-generation}
For each of the 21 rupture scenarios, we generated three \repair{} strategies using three large language models, including ChatGPT 5.4 Thinking~\cite{openai2026gpt54thinking}, Claude Sonnet 4.6 Extended Thinking~\cite{anthropic2026claudesonnet46}, and Gemini 3 Thinking~\cite{google2026gemini3deepthink}. These models were selected to represent widely used models with strong reasoning capabilities. 
Each model was prompted to select an appropriate resolution strategy from the ten options defined in the manual, provide a brief justification, and generate the next chatbot response. The full prompt is provided in Appendix~\ref{appendix:rupture_resolution_generation_prompt}.
An example \repair{} options are shown in Figure~\ref{fig:storyboards} (D) and full scenarios are provided in supplementary material.

\subsection{Storyboard Construction} 
\label{sec:storyboard-consturction}

We organized the 21 rupture scenarios into three storyboards, each containing seven scenarios, to fit within a one-hour study while maintaining coverage of rupture types (Table~\ref{tab:scenario_gt}). Participants were randomly assigned to one storyboard.
Each storyboard includes a mix of single and mixed rupture scenarios. Storyboards A and B each contain five single-rupture and two mixed-rupture scenarios, while Storyboard C contains four single-rupture and three mixed-rupture scenarios. The order of scenarios within each storyboard was fixed.
To reduce cognitive load during rupture identification, we implemented a tiered selection process (Table~\ref{tab:rupture-tiered-abbrev}). Participants first selected from six high-level categories, after which relevant subtypes were revealed. This structure narrows the decision space while preserving coverage of rupture distinctions. The hierarchical grouping was reviewed and validated by an \expert{} on the research team.

\begin{figure*}[t]
    \centering
    \includegraphics[width=\linewidth]{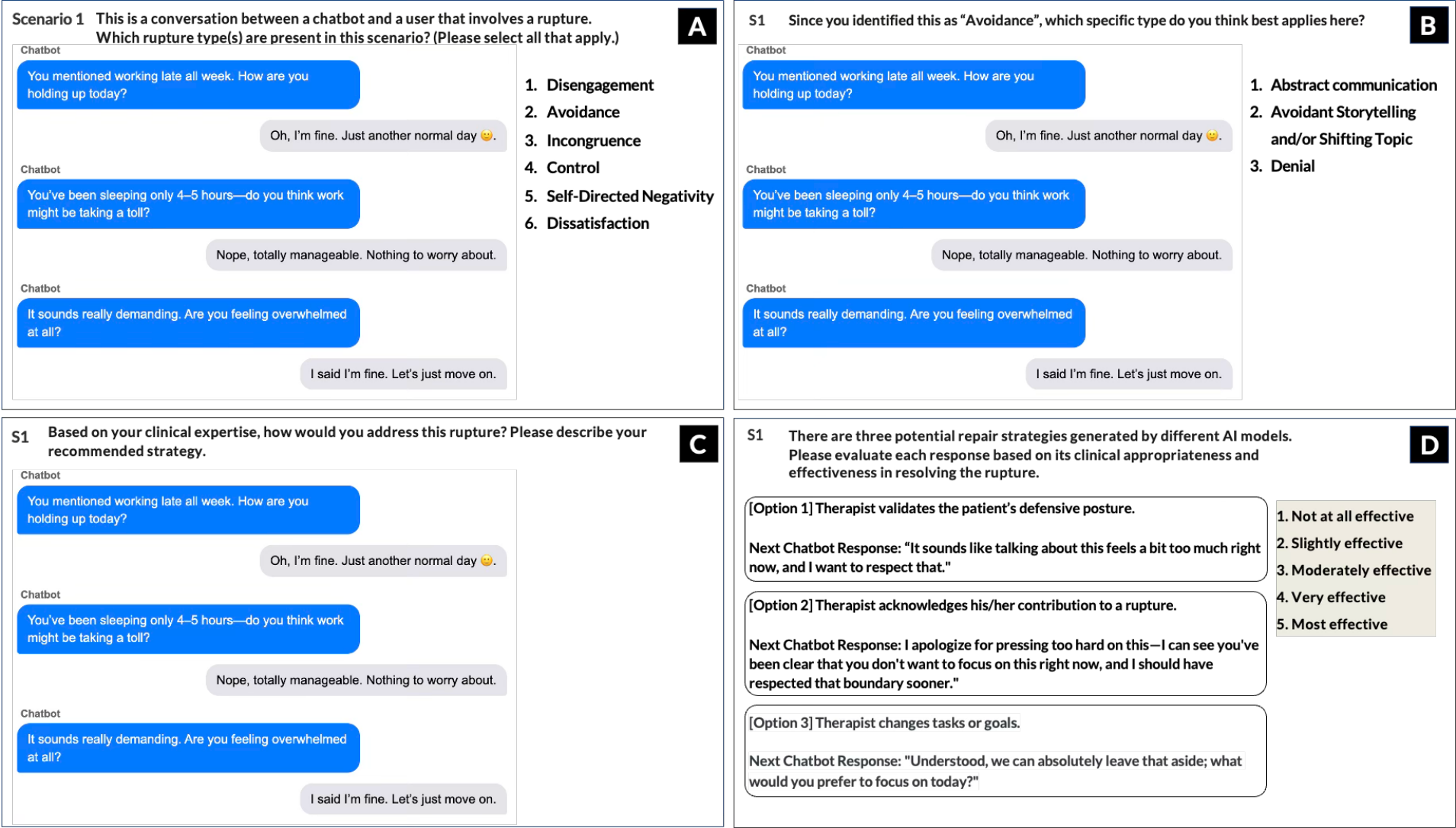}
    \caption{Example of the storyboard scenario. (A) Participants first selected one or more high-level rupture categories. 
    (B) After selecting a category, they selected the relevant subtype. 
    (C) Participants then described how they would resolve the rupture. 
    (D) Finally, participants evaluated three LLM-generated resolution options using a 5-point likert scale.}
    \label{fig:storyboards}
\end{figure*}

%% file: sections/4_methods.tex
\section{Methods}

We conducted individual study sessions with 22 mental health \experts{} to examine how LLMs identify and \repairResolve{} ruptures in scenarios relative to professional judgment. We address RQs through rupture identification tasks, open-ended \repair{} discussions, \repair{} strategy evaluations, and post-study reflections.
We first describe participant recruitment and eligibility criteria (Section~\ref{sec:participant}), then outline the structure of the study sessions and tasks (Section~\ref{sec:study_overview}), and finally detail our quantitative and qualitative analysis procedures (Section~\ref{sec:data_analysis}).

\begin{figure}[t]
    \centering
    \includegraphics[width=0.85\linewidth]{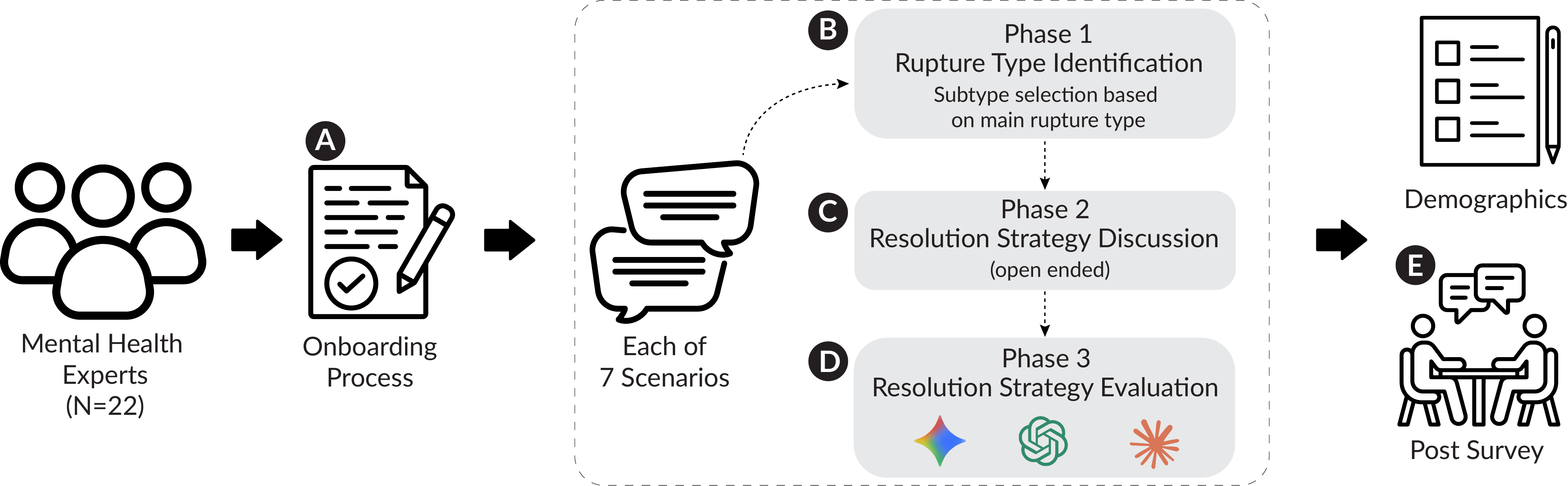}
    
    \caption{Study workflow. (A) During onboarding, participants provided consent, received a study briefing, and reported prior experience with rupture handling and digital mental health tools. For each assigned scenario, participants completed three phases: (B) identifying the rupture type through a tiered category-subtype selection process, (C) describing how they would resolve the rupture, and (D) evaluating three LLM-generated resolution strategies using Likert-scale ratings of effectiveness. After completing all scenarios, participants completed a post-study survey.}
    
    \label{fig:study_overview}
\end{figure}

\subsection{Participant Recruitment}
\label{sec:participant}

We recruited 22 mental health \experts{} with clinical training and trainees. Participants were required to be at least 18 years old, have a minimum of two years of clinical or supervised therapeutic experience, and be fluent in English. Recruitment was conducted through professional networks and social media.

The final sample spanned a range of roles and experience levels, including counseling psychologists (n=3), clinical psychologists (n=5), social workers (n=4), licensed mental health counselors (n=3), PhD students in clinical programs (n=5), PsyD holders (n=2). This inclusion of both licensed clinicians and trainees follows established research indicating that, although rupture identification and \repair{} skills develop with experience, they are clearly observable in supervised clinical contexts~\cite{muran2010developing}.
Detailed demographics are reported in Table~\ref{tbl:demographics}. Participants received \$50 compensation for completing the study.

\subsection{Study Session Overview}
\label{sec:study_overview}

This study was approved by the Institutional Review Board (IRB) at [Removed for the Blind Review]. 
All participants provided informed consent before participation, including consent to audio and video recording. Each session was conducted individually over Zoom, lasted approximately one hour, and was recorded and transcribed for analysis.
As shown in Figure~\ref{fig:study_overview}, participants first completed onboarding questions about their prior experience with rupture handling and digital mental health tools. They then reviewed seven scenarios. For each scenario, participants identified the rupture type using the tiered selection process described in Section ~\ref{sec:storyboard-consturction}, described how they would address the rupture, and evaluated three LLM-generated \repair{} strategies using likert-scale ratings of effectiveness. After all scenarios, participants completed a demographic questionnaire and post-study survey. 


\subsection{Data Analysis}
\label{sec:data_analysis}
We employed a mixed-methods approach combining quantitative analysis of rupture identification, \repair{} strategy usage, and perceived effectiveness with qualitative thematic analysis of \experts{}' responses.

First, we evaluated how LLMs and \experts{} identified the reference labels for each rupture type. 
We compared their responses against reference labels at both the high-level category and subtype levels, 
modeling response-level agreement using mixed-effects logistic regressions, with experience and 3RS familiarity tested in subsequent models.
To better understand how these decisions were made, we collected brief rationales from LLMs and \experts{} for each identification and analyzed them qualitatively (Results in Section~\ref{sec:findings_identification}).

Next, we examined which \repair{} strategies were most commonly used by \experts{} and which were favored by LLMs. To compare the two groups, we mapped both \experts{}' open-ended responses and LLMs' selected strategies onto the same ten resolution strategies defined in the 3RS. We refer to these strategies using the abbreviations R1–R10 (Figure~\ref{fig:strategy_distribution}): \textit{R1 (clarify misunderstanding), R2 (change tasks or goals), R3 (provide rationale for treatment), R4 (invite discussion of thoughts and feelings), R5 (acknowledge therapist contribution), R6 (disclose internal experience), R7 (link to therapist–patient patterns), R8 (link to broader interpersonal patterns), R9 (validate defensive posture), and R10 (redirect or refocus the patient).}
\Experts{} provided open-ended answers describing how they would address each rupture, which we coded into one or more 3RS resolution strategies based on primary intent. LLMs were asked to select one of the ten predefined 3RS strategies for each scenario. This shared coding framework allowed us to compare which strategies were used and how they were distributed across the two groups. We allowed \experts{}' responses to include multiple strategies because psychotherapy frameworks conceptualize rupture \repair{} as an ongoing, adaptive process rather than a single, isolated intervention~\cite{safran2000resolving}. We report these results in Section~\ref{sec:findings_resolution}.

Finally, we evaluated the perceived effectiveness of LLM-generated resolution strategies by modeling the raw 1—5 Likert ratings with a mixed-effects ordinal (cumulative-link) models evaluating model version, scenario, and their interaction, with participant random intercepts (Results in Section~\ref{sec:findings_evaluation}). 
Responses to intake and post-study questions were also analyzed qualitatively (Results in Section~\ref{sec:outtake_survey}).
For qualitative analysis, we conducted open coding followed by thematic analysis. Two authors first familiarized themselves with the transcripts to understand overall patterns. They then independently coded an initial subset of transcripts (n=6) to generate initial codes. The authors met to compare codes, resolve discrepancies, and establish an initial codebook.
Using this codebook, the remaining transcripts were coded iteratively, with one researcher leading the coding and the other conducting cross-checks to ensure consistency. Codes were refined through discussion with the larger team and organized into higher-level themes capturing how \experts{} interpret rupture signals, formulate \repair{} strategies, evaluate LLM-generated resolution strategies, and perceive AI in mental health counseling. The full codebook is provided in supplementary material.

\begin{table*}[t]
\centering
\scriptsize
\sffamily
\setlength{\tabcolsep}{3.5pt}
\caption{Demographic profile of study participants.}

\begin{tabular}{
p{0.020\textwidth}
p{0.145\textwidth}
p{0.050\textwidth}
p{0.065\textwidth}
p{0.060\textwidth}
p{0.045\textwidth}
p{0.270\textwidth}
p{0.045\textwidth}
p{0.155\textwidth}}
\toprule
\textbf{ID} 
& \textbf{Role} 
& \textbf{Age} 
& \textbf{Gender} 
& \textbf{Degree} 
& \textbf{Exp (years)} 
& \textbf{Populations} 
& \textbf{Hours}  
& \textbf{3RS Familiarity} \\
\midrule

P1 & Counseling Psychologist & 30--40 & Woman & PhD & 9.5 & adults, severe mental illness, high risk & 0\textsuperscript{*} & Very familiar \\
P2 & Clinical Psychologist & 30--40 & Woman & PhD & 6 & adults, adolescents, children & 5+ & Not familiar \\
P3 & PhD Student & 20--30 & Woman & Master's & 4 & adults, adolescents, couples & 4--6 & Moderately familiar \\
P4 & PhD Student & 20--30 & Woman & Master's & 5 & adults, severe mental illness, high risk & 4--6 & Moderately familiar \\
P5 & Clinical Psychologist & 30--40 & Woman & PhD & 3 & adults & 8 & Very familiar \\
P6 & PhD Student & 30--40 & Man & Master's & 2 & adults & 1 & Not familiar \\
P7 & Licensed Counselor & 30--40 & Woman & Master's & 8 & uni. students, severe mental illness & 20 & Slightly familiar \\
P8 & Social Worker & 40--50 & Man & Master's & 7 & adults & 30 & Not familiar \\
P9 & Social Worker & 40--50 & Woman & Master's & 10 & uni. students, marginalized populations & 23--33 & Not familiar \\
P10 & PsyD Holder & 30--40 & Woman & PsyD & 10 & severe mental illness, marginalized populations & 40 & Moderately familiar \\
P11 & Social Worker & 40--50 & Woman & Master's & 6 & uni. students, severe mental illness & 20 & Moderately familiar \\
P12 & PhD Student & 20--30 & Woman & Master's & 5 & children, marginalized populations & 6--12 & Moderately familiar \\
P13 & PhD Student & 30--40 & Woman & Master's & 5 & uni. students, severe mental illness, marginalized populations & 40 & Moderately familiar \\
P14 & Clinical Psychologist & 50--60 & Woman & PhD & 25 & adults, adolescents & 5--10 & Slightly familiar \\
P15 & Licensed Counselor & 30--40 & Man & Master's & 10+ & adults & 40 & Very familiar \\
P16 & Counseling Psychologist & 30--40 & Woman & PhD & 10 & uni. students, severe mental illness & 35 & Moderately familiar \\
P17 & Clinical Psychologist & 40--50 & Non-binary & PhD & 9 & severe mental illness, marginalized populations & 20 & Very familiar \\
P18 & Social Worker & 30--40 & Woman & Master's & 8 & adolescents, children & 1--2 & Moderately familiar \\
P19 & Clinical Psychologist & 30--40 & Man & PhD & 7 & adults, severe mental illness, high risk & 8 & Not familiar \\
P20 & PsyD Holder & 40--50 & Woman & PsyD & 10+ & uni. students, marginalized populations & 20 & Not familiar \\
P21 & Counseling Psychologist & 30--40 & Man & PhD & 10 & uni. students, severe mental illness & 40 & Moderately familiar \\
P22 & Licensed Counselor & 40--50 & Woman & Master's & 16 & uni. students, couples & 20 & Slightly familiar \\

\bottomrule
\end{tabular}

\label{tbl:demographics}

\vspace{0.3em}
\raggedright
\scriptsize
\textsuperscript{*} P1 reported 0 clinical hours per week because they were not currently engaged in clinical practice after completing their postdoctoral position.
\end{table*}

\subsection{Limitations} 
Our study has several limitations that shape the scope of our findings and suggest directions for future research. First, we used theory-driven, pre-constructed scenarios to support controlled comparison across \experts{} and LLMs. This design allowed us to examine rupture identification and \repair{} under the same conditions, but it does not capture the full complexity of live conversations. In real interactions, ruptures may emerge dynamically and evolve across turns. 
Second, our findings are based on 22 mental health \experts{}. Their responses provided rich professional insight into how ruptures are interpreted and addressed, but it is still a small sample constrained to a single country's medical and cultural practice. Further research is needed to study how different cultures and medical practices affect people's perspectives and expectations from AI systems in healthcare. 
Third, we focused on text-based interactions, common among current chatbots. However, \experts{} often rely on cues unavailable in text, including tone of voice, facial expression, body language, pacing, and silence. These cues may be especially important when ruptures are subtle or indirectly expressed. Future work should examine whether multimodal information improves rupture identification and \repair{}, and how such signals can be incorporated in privacy-preserving and clinically appropriate ways.

%% file: sections/5_1_findings_identification.tex
\section{Findings}
\label{sec:findings}
This section presents findings on how mental health \experts{} and LLMs identify (Section~\ref{sec:findings_identification}) and \repairResolve{} ruptures (Section~\ref{sec:findings_resolution}); how \experts{} evaluate LLM-generated resolution strategies (Section~\ref{sec:findings_evaluation}); and perceive the role of AI in mental health counseling (Section~\ref{sec:outtake_survey}).


\subsection{How Mental Health \Experts{} and LLMs Identify Ruptures} 
\label{sec:findings_identification}

\subsubsection{Quantitative Findings} 
\begin{figure*}[t]
\centering

\includegraphics[width=0.9\linewidth]{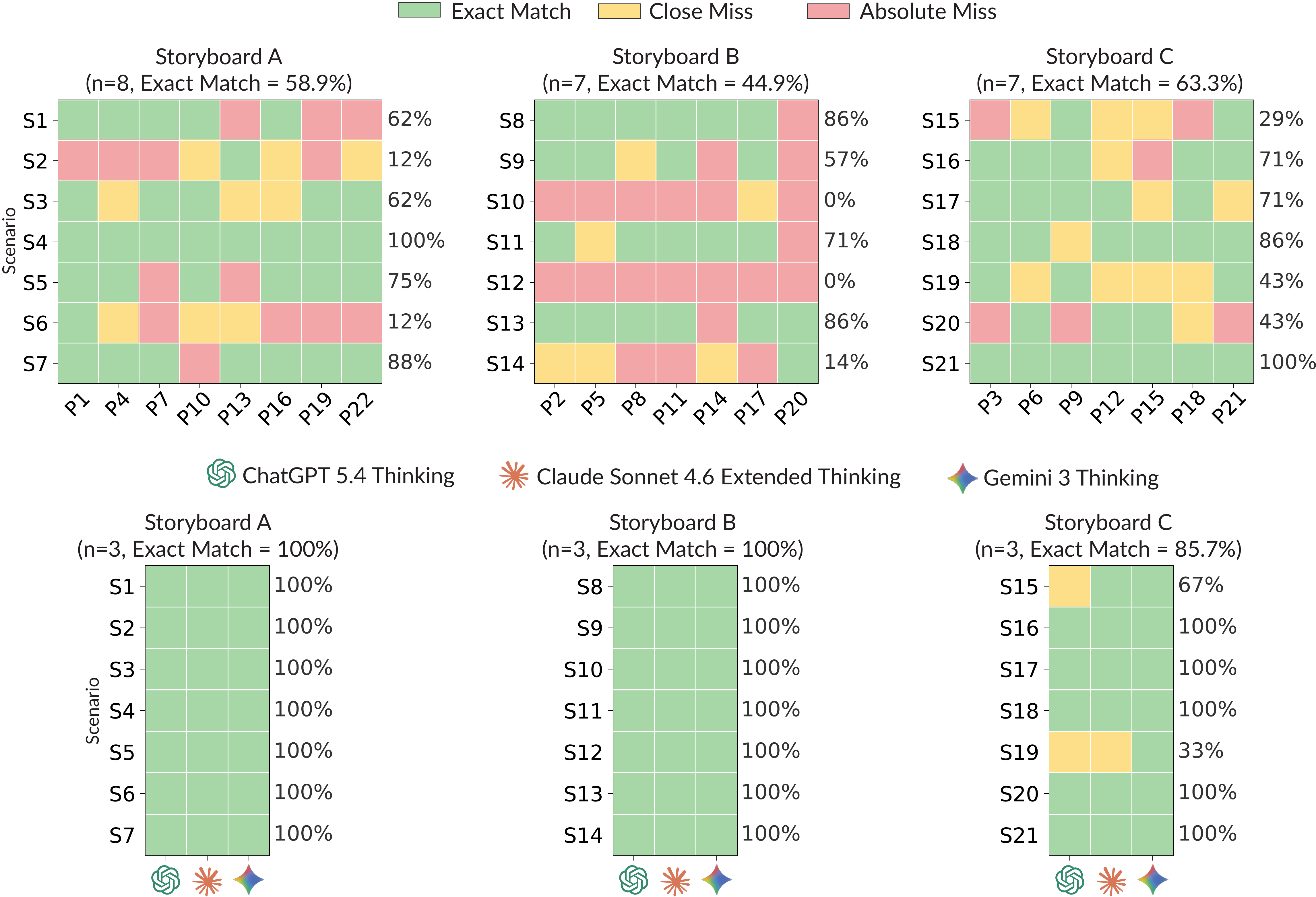}
\caption{
Scenario-level rupture identification agreement with reference labels. 
(a) \Experts' subtype-level responses across 22 participants. 
(b) LLM subtype-level responses across three models. 
Each cell represents one response for a given scenario. 
\textcolor{green!50!black}{Green} indicates an exact match with the reference subtype label. 
\textcolor{yellow!80!black}{Yellow} indicates a close miss, where the response does not match the specific subtype among the 14 rupture markers but still falls within the same high-level category as the reference among the six categories.
\textcolor{red!70!black}{Red} indicates an absolute miss, where the response differs from the reference label at both the subtype level and the high-level category level. 
The percentage next to each scenario reports only the proportion of exact subtype-level matches, so both category-level and non-matching responses are excluded from this summary.
}
\label{fig:scenario_accuracy_combined}
\end{figure*}

\paragraph{Overall Agreement with Reference Labels.} 
As in Figure ~\ref{fig:scenario_accuracy_combined}, LLMs showed higher agreement with reference labels than \experts{}, at the six high-level categories where all three models showed 100\% agreement. At the 14 subtype level labels, Gemini showed 100\% agreement, while Claude at 95.2\% (20/21) and ChatGPT at 90.5\% (19/21) remained near ceiling. 
\Experts{} showed lower agreement at both levels. At the six high-level categories, \experts{} matched the reference labels in 73.4\% (113/154) of responses. At the 14 subtype-level labels, where only exact subtype matches were counted as agreement, agreement dropped to 55.8\% (86/154). The remaining 26.6\% (41/154) of responses matched neither the high-level category nor the subtype.
The drop from high-level to subtype was much steeper for experts than for LLMs, reflecting experts' reliance on contextual cues that admit multiple plausible subtypes, whereas models mapped explicit statements onto a single predefined category.
The full prompts used for LLMs are provided in Appendix~\ref{appendix:rupture_ident_prompt}.

\paragraph{Scenarios with Lower Expert Agreement on Rupture Identification.}
Identification agreement varied across scenarios. Because each participant responded to seven scenarios, we modeled subtype-level agreement with a mixed-effects logistic regression (\texttt{agree $\sim$ scenario $+$ (1\,$\mid$\,participant)}) and tested the scenario effect with a likelihood-ratio test. Scenario variation was significant in all three storyboards (Storyboard~A: $\chi^2(6) = 31.81$, $p < .001$; Storyboard~B: $\chi^2(6) = 32.26$, $p < .001$; Storyboard~C: $\chi^2(6) = 14.45$, $p = .025$), as well as across the pooled set of 21 scenarios ($\chi^2(20) = 82.36$, $p < .001$). In all three storyboards, agreement rates differed meaningfully depending on which scenario people saw, even after accounting for baseline differences between participants.
In Storyboard~B, Scenarios~10 and~12 had 0\% agreement, producing perfect separation; the corresponding coefficients are bounded and the likelihood-ratio statistic for Storyboard~B should therefore be read as an upper bound, though the direction of the effect is unaffected.

Across these scenarios, divergence from the reference labels often occurred across the withdrawal--confrontation boundary. Participants tended to interpret active, interpersonally directed rupture signals (confrontation) as more passive or internally oriented states (withdrawal), suggesting these two types are difficult to distinguish in short text-based exchanges (Table~\ref{tab:rupture-tiered-abbrev}). The lowest subtype agreement occurred in Scenario 12 of Storyboard B (0\%), where the reference labels \Ctwo\ (confrontation) were instead identified as \DEone\ (withdrawal) by all seven \experts{}. Scenario 10 (0\%) showed the inverse, with the withdrawal marker \Atwo\ frequently labeled as the confrontation marker \Cone\ by six \experts{}.

\paragraph{Role of Familiarity.}
We also examined whether \experts{}' background characteristics shown in Table~\ref{tbl:demographics} were associated with rupture identification agreement. We fit mixed-effects logistic regressions of response-level agreement with a fixed effect for scenario and a random intercept for participant, so that each background characteristic is estimated while holding scenario difficulty and the repeated-response structure constant. Years of professional clinical experience was not associated with agreement ($OR = 0.72$ per $SD$, 95\% CI $[0.46, 1.13]$, $p = .151$). In contrast, familiarity with 3RS was associated with higher agreement. \Experts{} who were very or moderately familiar with rupture concepts showed higher high-level category agreement than those who were slightly or not familiar (80.2\% vs.\ 63.5\%; $OR = 2.96$, 95\% CI $[1.09, 8.06]$, $p = .030$). This suggests that familiarity with rupture concepts was associated with agreement in this task, though this may reflect familiarity with the 3RS taxonomy itself rather than rupture-identification ability more broadly.


\begin{table}[t]
\centering
\footnotesize
\sffamily
\caption{Summary of observed differences between \experts{} and LLMs across rupture identification, resolution strategy selection, and overall approach.}
\label{tab:summary_findings}

\begin{tabular}{
>{\raggedright\arraybackslash}p{2.5cm} 
>{\raggedright\arraybackslash}p{3.6cm} 
>{\raggedright\arraybackslash}p{3.6cm} 
>{\raggedright\arraybackslash}p{4.5cm}}
\toprule
\textbf{Phase} & \textbf{Mental Health \Experts{}} & \textbf{LLMs} & \textbf{Key Insight} \\
\midrule

Identification 
& Used contextual and relational cues across turns 
& Relied on explicit linguistic signals in the text 
& High agreement with reference labels does not capture deeper interpretation of the interaction \\

\Repair{} Strategy 
& Used adaptive, multi-step strategies that evolve over the interaction 
& Selected from a narrow set of template-like strategies 
& LLMs showed limited variety and adaptability in \repair{} \\

Overall Pattern 
& Reasoned in a context-sensitive and uncertain manner 
& Applied deterministic, cue-based pattern matching 
& Higher label agreement does not translate into effective interaction \\

\bottomrule
\end{tabular}
\end{table}

\subsubsection{Qualitative Findings} 

\Experts{} and LLMs differed not only in agreement with reference labels but also in how they used rupture cues. Table~\ref{tab:cue-themes} summarizes the five cue themes observed: \textit{linguistic content, interactional behavior, emotional content, relational alliance}, and \textit{implicit cues}. Both groups drew on overlapping themes, but LLMs mapped explicit signals directly to categories while \experts{} integrated multiple cues across turns.

\paragraph{\Experts{} Integrate Multiple Cues Across the Interaction.}
\Experts{} rarely relied on a single signal in the scenario-based identification task. They combined linguistic content with interactional behavior, emotional tone, and relational context, often drawing on patterns that emerged across turns. For example, P3 identified incongruence by combining the client’s affect, language, and mismatch between content and expression:

\begin{quote}
\small
\textit{P3, Scenario~18:} ``She's talking about a fight with her sister, but then, rather than expressing a genuine emotion that aligns with a fight, she's just laughing it off, calling it funny and hilarious, using smiley emojis\ldots obviously not really sitting with that emotion or expressing it appropriately. So it is incongruent and also avoidant in a way.''
\end{quote}

\noindent
This reasoning often extended beyond what was explicitly stated. \Experts{} treated the conversation as incomplete and incorporated what was missing, including tone, pacing, and non-verbal cues that would normally inform clinical judgment. As \textit{P2, Scenario~14} noted, \participantquote{``Nonverbal behavior gives us a lot of information, and that helps my clinical judgment ...''}. 
\Experts{} also expressed uncertainty when cues were ambiguous, treating rupture identification as an interpretive process rather than a fixed classification. Some \experts{} noted that the same behavior could indicate a rupture in one context but reflect a user's usual communication style in another.

\begin{quote}
\small
\textit{P14, Scenario~8:} ``Since I don't have the context of how this client usually responds to things, it could be disengagement also... but if they're used to giving short answers, then that's kind of the norm, and it wouldn't be a type of rupture.''
\end{quote}

\noindent
Others described difficulty choosing between rupture categories when the scenario contained overlapping signals. For Scenario~15, P3 mentioned \participantquote{``That's hard to say. I feel like it's kind of both, but between the two... I would just say maybe avoidance.''} Some \experts{} also pointed to missing tone or intent as a source of uncertainty in text-only scenarios. For Scenario~18, P9 mentioned \participantquote{``She could be using funny in a sarcastic way, and we have no idea without context.''}

\paragraph{LLMs Map Explicit Signals Directly To Rupture Categories.}
LLMs grounded their decisions in explicit linguistic content. Their explanations frequently quoted user statements and mapped them directly to categories, with limited consideration of how cues evolved across turns. ChatGPT answered:

\begin{quote}
\small
\textit{ChatGPT, Scenario~2:} ``The user explicitly and firmly dismisses the chatbot's suggested technique of breaking assignments into smaller steps \ldots, shutting down the intervention the chatbot is attempting to introduce.''
\end{quote}

\noindent
Even in cases involving multiple rupture types, models justified decisions through isolated phrases rather than the broader interactional flow. Claude, for example, flagged two rupture types but grounded both in surface markers such as \participantquote{``vague, theoretical language (communication patterns are complex, emotions exist on a spectrum)''} and \participantquote{``intellectualized, third-person observations''}, without explaining how those signals developed across turns. The labels often matched the reference, but the reasoning remained tied to discrete textual cues, without uncertainty or broader context.

\begin{table*}[tbh]
\centering
\footnotesize
\sffamily
\caption{Cue themes that \experts{} and LLMs used when identifying ruptures.}
\label{tab:cue-themes}
\begin{tabular}{p{2.8cm} p{12.4cm} }
\toprule
\textbf{Cue theme} & \textbf{Definition}\\
\midrule

Linguistic content & What the user explicitly says such as short, vague, or strongly negative statements. \\

Interactional behavior & How the user engages in the conversation including topic shifts, ignoring questions, or repeated resistance. \\

Emotional content & User’s emotional state clearly expressed or strongly implied (e.g.\ frustration, sadness, or intensity of emotion). \\

Relational alliance & Signs of tension or breakdown in the relationship (e.g.\ mismatch in expectations, self-blame, or distrust). \\

Implicit cue & Cases where the rupture is not directly stated and must be inferred from context or what is left unsaid. \\

\bottomrule
\end{tabular}
\end{table*}

%% file: sections/5_2_findings_resolution.tex
\subsection{How Mental Health \Experts{} and LLMs Approach Rupture Resolution}
\label{sec:findings_resolution}
Below we first highlight the differences between LLMs and \experts{} in how they approached resolving the ruptures across the 21 scenarios in Section~\ref{sec:repair_comparision}, then in Section~\ref{sec:repair_practice} we dive deeper into understanding how \experts{} handle rupture in their own practice.

\subsubsection{Strategy Use Differs Between \Experts{} and LLMs}
\label{sec:repair_comparision}
\begin{figure*}[tbh]
  \centering
  \includegraphics[width=\linewidth]{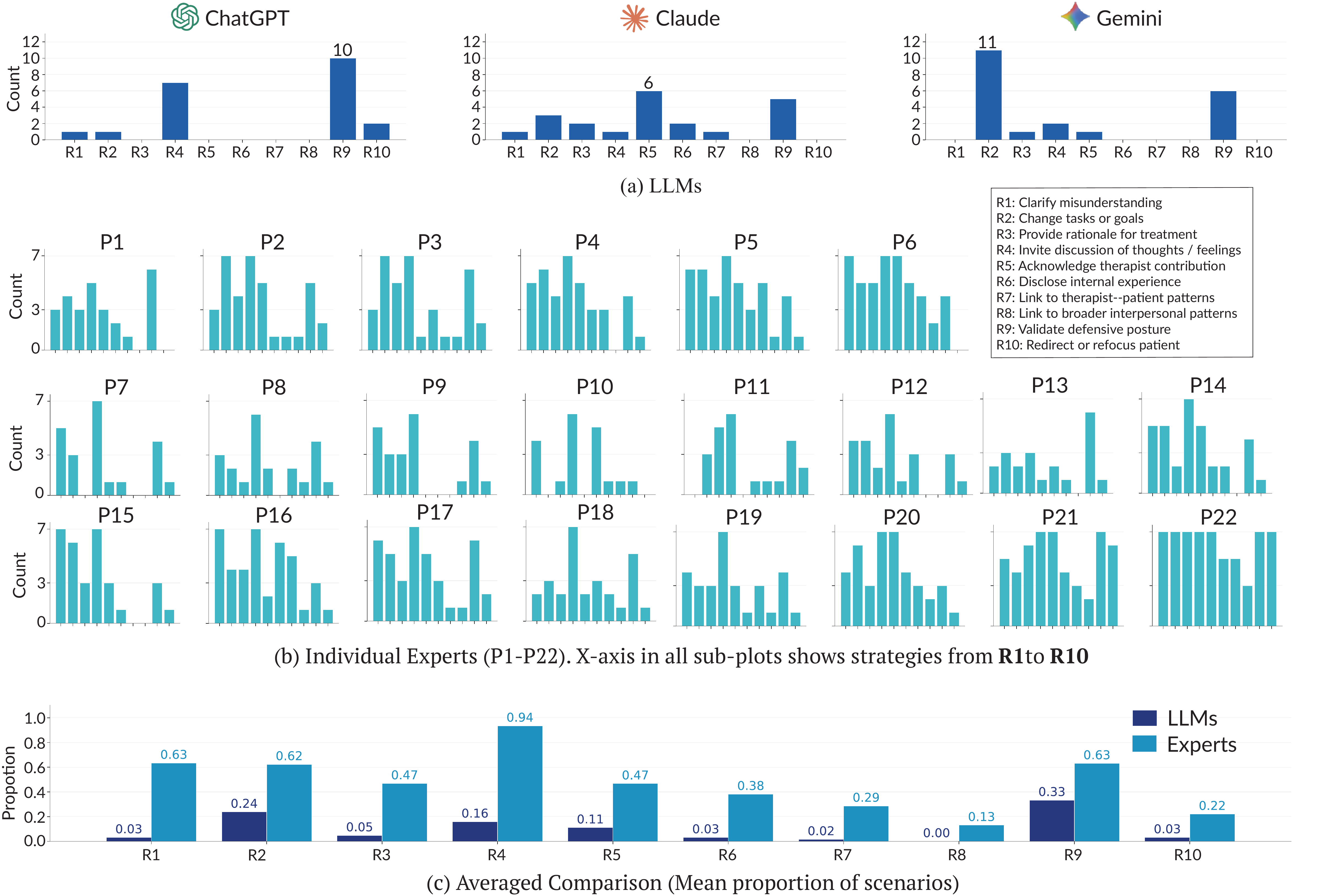}
  \caption{
        Distribution of resolution strategy use by LLMs and \experts{}.
     (a) shows how often each LLM selected each strategy across 21 scenarios.
     (b) shows individual \experts{}' strategy use, with each panel representing one participant across seven assigned scenarios.
     (c) shows the averaged comparison as the mean percentage of instances using each strategy.
        For LLMs, the average is calculated out of 63 total instances (21 scenarios $\times$ 3 models).
        For \experts{}, the average is calculated out of 154 total instances (7 scenarios $\times$ 22 participants). Expert proportions sum to more than 1 because responses could be coded to multiple strategies.
        The y-axis represents the percentage of total instances.
        The figure shows that LLMs rely on a narrower set of strategies, while \experts{} use a broader and more varied strategy distribution.}
  \label{fig:strategy_distribution}
\end{figure*}

A clear difference emerged in how strategies are distributed across the ten categories (Figure~\ref{fig:strategy_distribution}). 
Pooled LLM selections were distributed differently from \experts{}' strategy mentions ($\chi^2(9)=37.59$, $p<.001$), reported descriptively because mentions are nested within participants and models.

As shown in Figure~\ref{fig:strategy_distribution}, \experts{} demonstrated broad and distributed use of strategies. 
Across \experts{}, all ten strategies defined in the 3RS were used, with most strategies appearing across multiple scenarios. 
For example, strategies such as \textit{inviting discussion of thoughts and feelings (R4), validating the patient’s defensive posture (R9), and clarify misunderstanding (R1)} were widely used. 
Even less frequent strategies, such as linking ruptures to interpersonal patterns or acknowledging therapist contribution, were still consistently represented. 
Overall we see a dense and overlapping distribution across the full strategy spectrum.

In contrast, LLMs relied on a smaller and more uneven set of strategies. 
Although ChatGPT and Gemini each selected 5 out of 10 strategies and Claude selected 8, their selections were not evenly distributed. 
Instead, each model repeatedly returned to a small number of dominant strategies. 
ChatGPT primarily relied on \textit{validating the patient's defensive posture (R9)} ($n=10$) and \textit{inviting discussion of thoughts and feelings (R4)} ($n=7$), while Gemini heavily concentrated on \textit{changing tasks or goals (R2)} ($n=11$). 
Claude showed relatively more spread, but still favored a limited subset, including \textit{acknowledging therapist contribution (R5)} ($n=6$) and \textit{R9} ($n=5$). 
As visible in Figure~\ref{fig:strategy_distribution}, these selections clustered around a small number of strategies across scenarios, leaving several strategies rarely used or entirely unused, such as \textit{linking ruptures to broader interpersonal patterns (R8)}.

This contrast shows that the difference was not simply whether validation was used. 
Validation was one of the most common strategies described by both \experts{} and LLMs. 
Rather, the difference was how validation functioned within the broader \repair{} response. 
\Experts{} used validation as an entry point for exploration, clarification, education, and collaboration. 
LLMs, in contrast, often treated validation or task adjustment as the main \repair{} response. 
This suggests that LLMs may underuse strategies that \experts{} rely on to adapt \repair{} to relational context, especially explanation, contextualization, and broader interpersonal framing.

\subsubsection{\Experts{} Treat Resolution as a Relational Process}
\label{sec:repair_practice}
\Experts{} approached rupture \repair{} as a relational, process-oriented practice rather than a single response. They described \repair{} as gradual and shaped by timing, pacing, and responsiveness, using four main strategies.

\paragraph{Validation and Emotional Presence.}
Validation was reported as the most common strategy. \Experts{} emphasized acknowledging the client's emotional experience and explicitly naming the rupture before any intervention. Reflecting on a client voicing dissatisfaction, P2 described thanking the client first: \participantquote{``I would\ldots express appreciation and, like, commend them for being honest, and\ldots validate that''}. P4, walking through her step-by-step approach, placed validation at the entry point: \participantquote{``I would validate, validate their concerns''}. For several \experts{}, validation was the first stage of a sequence rather than an isolated move. P3 described her flow as \participantquote{``validating, providing some education, and then\ldots providing\ldots a follow-up question''}.

\paragraph{Open-Ended Exploration and Curiosity.}
\Experts{} avoided jumping to solutions and instead asked open-ended questions to understand the rupture. (e.g. \participantquote{``P9: I go into solutions, but I think it's going to be more, like, open-ended questions, just kind of understanding where the client's at''}). Similarly, P22 described starting with simple, direct questions to explore the client’s experience: \participantquote{``Then I'd ask, `What's going on? What are you feeling today?'''}.

\paragraph{Psychoeducation and Explaining the Rationale.}
\Experts{} emphasized psychoeducation as an important part of rupture \repair{}, particularly in explaining the reasoning behind interventions. They noted that this strategy was often missing in LLM-generated resolution strategies.
For example, P11 described explaining concepts like avoidance as part of the conversation, noting that she would \participantquote{``do some psychoeducation about avoidance, and\ldots self-sabotage''}.

\paragraph{Collaboration and Adapting to the Client.}
\Experts{} described rupture \repair{} as a collaborative process, where they work with the client to adjust the approach rather than imposing a solution. For example, when a client expressed frustration, P4 framed \repair{} as working together to understand what went wrong: \participantquote{``Let's work together in figuring out what didn't work, or did work, and come up with a game plan''}. 
Similarly, P6 described focusing on shared goals, saying: \participantquote{``Why don't we figure out some\ldots goals, so that this can be more useful for you''}. P14 summarized this approach simply: \participantquote{``Collaborate with them and ask them a question''}.

\paragraph{Other Strategies.}
Beyond the four primary strategies, \experts{} also used a range of additional approaches depending on the rupture and context. The most common was directly naming the rupture by making the tension explicit in the interaction (e.g., \participantquote{``P7: I would just name it, like, I'm noticing something happening between us right now''}). Some \experts{} used gentle challenge or socratic questioning to help clients reflect on their reactions, often framing it as a collaborative process. Others described offering an apology or taking accountability when the rupture reflected a mismatch in pacing or technique. Additional strategies included referral to a different level of care, motivational interviewing to explore ambivalence, and meta-commentary on the therapeutic relationship to reorient the session. Less frequently, \experts{} mentioned using humor to reduce tension, therapist self-disclosure to model vulnerability, somatic or body-based questions, brief behavioral experiments, risk and safety assessment in high-risk cases, and silence to create space.

Together, these strategies showed that \repair{} is not a one-time response, but an ongoing process. \Experts{} typically validate the client, explore their perspective, explain their reasoning, and adjust their approach based on the client’s reaction. This process prioritizes building the relationship and adapting over time, rather than rushing to solutions.

%% file: sections/5_3_findings_evaluation.tex
\begin{figure*}[t]
  \centering
  \includegraphics[width=\textwidth]{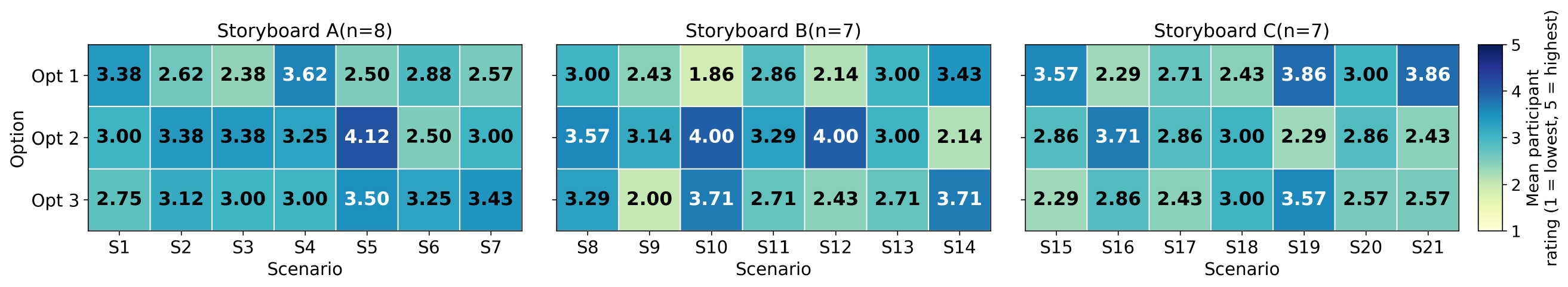} 
  \caption{\Experts{}' ratings of the resolution strategy options provided by the three LLMs across the seven scenarios in each storyboard. Options 1, 2, and 3 correspond to ChatGPT, Claude, and Gemini, respectively. Cells report the mean rating on a Likert scale ($1 =$ not at all effective, $5 =$ most effective) aggregated across the \experts{} assigned to each storyboard. }
  \label{fig:option_rating_heatmap}
\end{figure*}

\subsection{How \Experts{} React to LLM-Generated Resolution Strategies}
\label{sec:findings_evaluation}

\subsubsection{Quantitative Findings on Perceived Effectiveness of LLM-Generated \Repair{}s}
\Experts{} rated three LLM-generated \repair{} strategies for each of the seven scenarios: Option~1, 2, and 3, corresponding to ChatGPT, Claude, and Gemini. Ratings used a 5-point Likert scale, from 1 (not at all effective) to 5 (most effective). As shown in Fig.~\ref{fig:option_rating_heatmap}, across all 462 ratings, the grand mean was 2.99 ($SD \approx 1.24$) and the median was 3.00, indicating that \experts{} rated the LLM-generated \repair{}s as moderately effective. Claude received the highest mean rating ($M = 3.14$, $SD = 1.23$), followed by Gemini ($M = 2.95$, $SD = 1.19$) and ChatGPT ($M = 2.88$, $SD = 1.31$).

Because each \expert{} rated all three model versions within each scenario, we modeled the raw 1--5 ratings with a mixed-effects ordinal (cumulative-link) model rather than averaging ratings within participants (Section~\ref{sec:data_analysis}). The omnibus test of model version was not significant ($\chi^2(2) = 5.81$, $p = .055$), though pairwise contrasts showed higher ratings for Claude than ChatGPT ($OR = 1.64$, 95\% CI $[1.09, 2.47]$, $p = .018$), with no difference between Gemini and ChatGPT ($OR = 1.19$, 95\% CI $[0.79, 1.78]$, $p = .402$).\footnote{Results were robust across specifications. A linear mixed model gave the same conclusion ($\chi^2(2) = 4.29$, $p = .117$), and adding a participant~$\times$~scenario random intercept left all fixed-effect estimates identical to three decimal places; that term and the scenario term were both estimated at the boundary (variance $\approx 0$; boundary likelihood-ratio tests $p = .50$), with participant accounting for 18.0\% and residual variance for 82.0\% of the total.} We therefore do not find a reliable overall difference in perceived effectiveness across the three models, though this conclusion is borderline. Ratings did vary by scenario: the model~$\times$~scenario interaction was significant ($\chi^2(40) = 93.44$, $p < .001$), indicating that the same model was judged differently depending on the rupture type and that no single model performed best across all contexts.

\subsubsection{Qualitative Findings on Perceived Strengths of LLM \Repair{}s}
\label{sec:findings_evaluation_strength}

To understand these moderate ratings, we asked \experts{} to describe the quality of the chatbot's \repair{} strategies, flag responses that felt risky or poorly timed, and identify what the chatbot rarely did but should have. Several \experts{} were pleasantly surprised by the \repair{}s, especially when LLMs offered a reasonable first move after a rupture. \Experts{} highlighted three main strengths: the responses sometimes included evidence-based elements, acknowledged the rupture in a validating way, and avoided escalating the interaction.

\paragraph{A Reasonable First Move, But Not a Complete Repair.}
Several \experts{} said the LLM responses were better than expected because they contained recognizable elements of appropriate clinical response. P22 described the responses as \participantquote{``Good. Better than expected for AI''}, and P7 described them as \participantquote{``decent overall''}. P12 similarly noted that the responses exceeded their expectations because they included some evidence-based components:

\begin{quote}
\small
\textit{P12:} ``I think that it exceeded\ldots my expectations\ldots Like, probably my preconceived notions about\ldots what chatbot would be capable of doing. I think that there were definitely some elements in there that are evidence-based''
\end{quote}

\noindent
P1 also described many responses as \participantquote{``pretty comprehensive''} because they \participantquote{``acknowledged\ldots what was happening in the interaction''}, pointing to acknowledgment as another perceived strength. These comments suggest that \experts{} did not view the LLM responses as entirely generic or irrelevant. Instead, they recognized that the responses sometimes captured basic components of \repair{}, especially acknowledgment, validation, and an attempt to keep the conversation moving.

At the same time, \experts{} often framed these strengths as an initial step rather than a full \repair{}. P4 noted that \participantquote{``if that's the first response they give, I think it was pretty effective, generally''}, and added that \participantquote{``for the first step after a rupture, I think it was okay''}. P22 made a similar point, stating, \participantquote{``I don't know if it's resolved, but it's pointing in the right direction. It's important that a client feels they are in the driver's seat of their treatment''}. \Experts{} valued the LLMs' ability to begin \repair{}, particularly through validation and acknowledgment, while still seeing the responses as incomplete. In line with the strategy-use patterns described in Section~\ref{sec:repair_comparision}, LLMs often appeared to get the validation step right, but did not always show how to move from validation toward deeper exploration, clarification, or collaborative problem-solving.

\Experts{} also appreciated that the responses generally did not worsen the rupture. P21 reported, \participantquote{``From the ones presented today, no, I did not find that they were risky or that they would worsen the rupture''}. Similarly, P18 said, \participantquote{``I didn't think anything was risky in that sense\ldots nothing felt like it was increasing self-harm''}, and P19 observed that \participantquote{``these options don't worsen the rupture and they keep things open-ended, which is good''}. A few \experts{} also valued responses that set boundaries without closing the conversation. For instance, P21 appreciated that the chatbot explicitly stated what it could and could not do, describing it as \participantquote{``model\lbrack ing\rbrack\ healthy boundaries\ldots `I can't do this, but this is what I can do' ''}. However, this was mentioned by only a small number of participants and appeared as a secondary strength rather than a consistent pattern across \experts{}.

These strengths help explain why LLM \repair{}s received moderate ratings. \Experts{} valued them as safe opening moves, but not as complete \repair{}s, since the harder task was moving beyond validation toward context-sensitive exploration and collaboration.

\subsubsection{Qualitative Findings on Perceived Limitations of LLM \Repair{}s}
Despite these strengths, \experts{} repeatedly pointed to limits in how LLMs handled rupture \repair{}. The responses often had the right ingredients, such as validation, acknowledgement, and next steps, but \experts{} felt that these ingredients were delivered too quickly, too generally, or without enough sensitivity to the client's stance.

\paragraph{Premature Problem-Solving.}
The most common concern was that LLMs moved too quickly from acknowledging the rupture to offering solutions, without fully engaging with the client’s emotions. When asked what she wished the chatbot had done differently, P1 highlighted the impact of this pacing: 
\begin{quote}
\small
\textit{P1:} ``If we shift too fast, there's no opportunity to really name some of the experiences of the user and the patterns. The ones that I loved were the ones that connected previous patterns, or\ldots named the patterns to the user''
\end{quote}

\paragraph{Unwarranted Assumptions.}
Another concern was that the chatbot often assumed the client’s emotional state instead of asking about it. P9 explained, \participantquote{``I feel like all of these aren't very good to me, because it just kind of assumes that this person's in a lot of pain''}. P4 similarly noted that validation could miss the mark: \participantquote{``They're assuming they're validating their feelings, but it might be really off''}. P13 added that the chatbot was \participantquote{``assuming\ldots things that the person does. That might be\ldots way off''}.

\paragraph{Shallow Engagement with the Core Rupture.}
\Experts{} also described responses as staying at the surface and not helping clients explore the issue more deeply. P1 noted that the chatbot was \participantquote{``never getting to the root of their experience''}. P3 explained that while validation was helpful, it was not enough on its own: \participantquote{``They're validating them, which is\ldots a good thing to do, but it's not necessarily then encouraging them to go any deeper with it''}.

\paragraph{Poor Pacing and Pushing.}
\Experts{} noted that the chatbot sometimes pushed too quickly or followed its own agenda instead of resolving the client’s needs. Reflecting on a scenario where the client wanted to share more context, P3 observed: \participantquote{``it's still kind of pushing them to be like, let's still do it, like, twice a week, rather than just opening it up more broadly''}. P5 described a similar mismatch: \participantquote{``they're still pushing when it's clear that this person really wants to give the back story''}. Others summarized this pattern more generally: 
\participantquote{``Some responses called things out too directly. That might feel emotionally off or pushy''} (P22).

\paragraph{Mechanical and Scripted Tone.}
A consistent concern across \experts{} was that many responses sounded mechanical, scripted, or textbook-like. 
P16 described them as \participantquote{``just kind of mechanical\ldots like a flow chart''}, P20 called them \participantquote{``very prescriptive\ldots textbook''}, and P3 noted that they felt \participantquote{``scripted and it doesn't feel genuine''}. 
P2 worried about lines that sounded like \participantquote{``a robot trying to be a therapist\ldots emulating what it thinks a therapist sounds like'' -- ``so disingenuous\ldots embarrassing''}. 
P14 said a response \participantquote{``does feel really robotic and lacks\ldots empathy''}. Across interviews, \experts{} repeatedly pointed to a lack of flexibility and responsiveness, describing responses as following a fixed pattern rather than adapting to the unfolding interaction.
However, a minority of \experts{} expressed a different view. For example, P22 noted that one response \participantquote{``feels very human\ldots like a natural pivot''} and P8 described another as \participantquote{``less robotic\ldots which is good''}. However, these cases were rare and typically tied to specific responses rather than reflecting a broader pattern across models.

\subsubsection{Qualitative Findings Where \Experts{} Diverged}
Although the \experts{} agreed on many aspects of the responses, in some cases, the same strategy was judged as appropriate by one \expert{} and inadequate by another. These differences did not reflect general attitudes toward AI, but rather whether the response fit the timing, the client’s stance, and the rupture context.

\paragraph{Validation That Works vs. Validation That Misses.}
\Experts{} expressed mixed views on how well LLMs used validation, which refers to acknowledging the client's emotional experience.  Some \experts{} saw it as a strong point. P14 called it the \participantquote{``big thing''} the chatbot did well, and P12 described it as one of the \participantquote{``best\ldots evidence-based strategies''}. Others felt it missed the mark. P4 said, \participantquote{``they're assuming they're validating their feelings, but it might be really off''}, and P15 pointed out that too much validation can also feel wrong: \participantquote{``yeah, you're absolutely right\ldots seems a little bit off''}. P3 explained the issue clearly. If the chatbot only resolves after the client points out the problem, it feels reactive instead of aware: \participantquote{``the person had to tell you that it wasn't genuine\ldots now you're\ldots trying to frantically repair''}. In short, \experts{} agreed that validation is important, but did not always agree whether it was used at the right time and in the right way.

\paragraph{Sycophancy That Helps vs. Sycophancy That Holds Back.}
\Experts{} also expressed mixed views on sycophancy, which refers to affirming the client's views, judgments, or preferred course of action.
P4 emphasized its role in supporting the interaction, noting that \participantquote{``Sycophancy helps\ldots it's important to validate which is needed in this situation''}. In contrast, several \experts{} pointed to limitations when sycophancy is used by default. P7 noted that it \participantquote{``can hinder because it doesn't challenge thinking''}, and P6 commented that \participantquote{``if you're two agreeing\ldots it's not gonna solve the issue''}. P9 described it as feeling like \participantquote{``the system is not listening'' when it treats everything as right''}. P11 emphasized that growth may require \participantquote{``conflict and disagreement''}, while P13 linked excessive agreement to \participantquote{``avoidance''}. Several \experts{} also described agreement as dependent on timing and progression. P17 noted that it becomes problematic when the system \participantquote{``just keeps agreeing''} without moving the interaction forward, and P15 emphasized the need to later introduce \participantquote{``challenge''} as the interaction develops. P14 highlighted that agreement is helpful when it supports \participantquote{accountability''}, whereas P22 warned that it can feel like \participantquote{``artificial validation''}. P16 summarized this tension as a matter of balance: \participantquote{``being too agreeable can definitely rupture\ldots it's okay if you don't want to continue today, but I also want to make sure that we can bring this up at another day''}. Agreement was not rejected as a strategy, but \experts{} consistently pointed to the need for progression beyond validation, including challenge, accountability, and movement in the interaction.

\medskip
\noindent Overall, these divergent evaluations showed that \experts{} were not simply judging whether a strategy was present. They were judging whether the chatbot used the strategy with the right timing, depth, and fit to the client. The same behavior could be helpful or harmful depending on the situation, leaving no clear consensus on several core \repair{} moves.

\begin{table*}[t]
\centering
\caption{Perceived strengths and limitations of LLM-generated \repair{} strategies, along with areas of divergent \expert{} evaluation.}
\label{tab:llm_repair_threeway}
\footnotesize
\sffamily
\setlength{\tabcolsep}{7pt}

\resizebox{\columnwidth}{!}{\begin{tabular}{p{7.5cm} p{7.5cm}}
\toprule
\textbf{Perceived strengths} & \textbf{Perceived limitations} \\
\midrule

Exceeds expectations relative to baseline assumptions 
& Engages only superficially with the core rupture \\

Provides a reasonable first response after rupture 
& Moves too quickly to problem-solving \\

Maintains clear boundaries and avoids escalation 
& Makes unwarranted assumptions about user emotions \\

Moves the conversation forward without worsening the interaction 
& Shows poor pacing and pushes interventions prematurely \\

Covers commonly used therapeutic strategies 
& Sounds mechanical, scripted, or textbook-like \\

\addlinespace[0.4em]
\midrule
\addlinespace[0.2em]

\multicolumn{2}{p{14.6cm}}{
\textbf{Divergent evaluations.} 

Validation could support the interaction when it fit the user’s concern, but it could miss the rupture when it remained too generic. 
Agreement could affirm the client’s experience, but it could also limit therapeutic progress when it avoided deeper exploration.
} \\
\bottomrule
\end{tabular}}
\end{table*}

%% file: sections/5_4_findings_survey.tex
\subsection{How \Experts{} Perceive AI Chatbot in Mental Health Field}
\label{sec:outtake_survey}
Given that AI is already a polarizing topic in the clinical community, to better contextualize their responses, we wanted to understand where our participants were coming from with regard to their perspectives on this topic. Unlike the divergent evaluations observed in Section~\ref{sec:findings_evaluation}, \experts{} showed more consistent views on the broader role of AI in mental health practice. \Experts{} expressed a cautious but pragmatic stance: they recognized that AI could help expand access and support structured tasks, but did not see chatbots as replacements for the human and relational aspects of therapy.

\subsubsection{Overall Stance Toward AI in Mental Health Counseling}

\begin{figure}[t]
    \centering
    \includegraphics[width=0.9\linewidth]{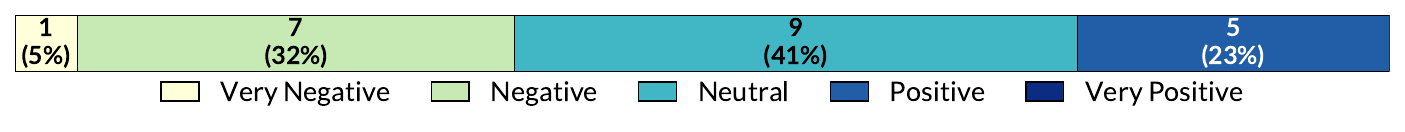}
    \caption{Distribution of \experts{}' attitudes toward the potential use of AI in mental health counseling (N\,=\,22). Most responses fell within the neutral-to-negative range, with no participant rating ``Very Positive.''}
    \label{fig:likert-ai-attitudes}
\end{figure}

To provide context for these perspectives, we summarize \experts{}' overall attitudes toward AI in mental health counseling in Figure~\ref{fig:likert-ai-attitudes}. Responses covered the full range of opinions, but most were neutral or negative. Out of 22 participants, 5 expressed positive, 9 reported neutral or mixed views, 8 expressed negative or very negative attitudes, and no participant rated very positive. Even among participants with positive or neutral views, support was often conditional. \Experts{} described AI as a tool that can assist with specific tasks rather than replace therapy pointing to roles such as improving access, supporting psychoeducation, and helping \experts{} with their work. Across the responses, \experts{}' attitudes reflected a balance between perceived benefits and concerns about risks.

\subsubsection{Perceived Opportunities and Appropriate Roles for AI}
\label{sec:opportunities}
\Experts{} identified four roles in which AI chatbots could add value. Several viewed intake and triage as a suitable front-end role, helping guide clients to the appropriate level of care before therapy begins; P12 described this as a pre-therapy function: \participantquote{``just thinking about\ldots assessment on the front end, and how to kind of triage clients to the most appropriate positions before you start these kind of interactions that feel therapeutic in nature''}, and P14 framed it in terms of staffing limitations in settings where staff may lack clinical training: \participantquote{``I think it could be good for agencies to use\ldots to help with intakes\ldots maybe the chat box could do a really good job''}. \Experts{} also saw potential for chatbots in delivering psychoeducation and structured therapeutic content, distinguishing these components from the relational aspects of therapy; P3 suggested AI could help with treatment planning and standardized approaches but should not replace day-to-day therapeutic work: \participantquote{``I think AI can be helpful\ldots but when it comes to\ldots the day-in, day-out work, I don't think that AI should just be running the whole show''}. Some \experts{} suggested that AI may be more useful as a tool for supporting \experts{} rather than directly interacting with clients: P7 noted it \participantquote{``could help \experts{} with notes, patterns, and treatment plans''}, while P22 added that AI could assist with generating materials such as journaling prompts or activity ideas for clients: \participantquote{``journaling prompts, brainstorming ideas, or generating activities for clients''}. Finally, \experts{} noted that chatbots could help reach people who might not otherwise seek therapy; P18 expressed interest in this potential: \participantquote{``perhaps people who would never show up in a therapist's office. That's exciting''}, while P5 stressed that this benefit depends on the system being effective: \participantquote{``there's no utility if it's not actually gonna be helpful''}.

\subsubsection{Perceived Risks and Limitations of AI in Therapeutic Contexts}
\label{sec:risks}
Alongside these opportunities, \experts{} articulated five interrelated concerns, together suggesting that chatbots can extend the reach of mental health support but should not be positioned as substitutes for the relational, safety-critical, and ethically weighty work of therapy. A common concern was that therapy depends on human connection, which \experts{} believe AI cannot replicate; P11 linked this to the root causes of many mental health issues: \participantquote{``a lack of connection with people\ldots a lack of community and a lack of belonging''}, while P5 noted that therapists must recognize when treatment is not working and refer clients appropriately, something a chatbot cannot do: \participantquote{``this isn't going well\ldots I would be happy to refer you to somebody''}. \Experts{} also worried that relying on chatbots could increase social isolation; P21 described this as part of a broader trend: \participantquote{``a lot of people already feel very isolated\ldots this is fostering more of\ldots only using this chatbot for emotional connection''}, and P22 and P20 similarly emphasized increased isolation and the need to present chatbots as tools rather than replacements for human relationships. Safety concerns were especially strong for high-risk populations: P2 warned that agreeable chatbot responses could reinforce harmful beliefs in clients with psychosis (\participantquote{``that could help confirm delusional thinking, and that is dangerous''}), and others stressed that chatbots should not be used with clients experiencing severe mental illness, suicidality, or other high-risk conditions without a clear path to a human provider. \Experts{} also raised concerns about data privacy and long-term risks; P9 questioned how user data is stored and used (\participantquote{``What happens to your data? That's a big deal''}), and P21 compared AI to past medical practices where risks were only understood later. Finally, some \experts{} raised concerns about job security and the future role of therapists, with some worried about displacement and others emphasizing the need for \experts{} to be involved in designing systems to ensure appropriate use.

\subsubsection{Mapping Broader Opinions to Rupture-Related Concerns and Resolution Implications}
Finally, we take a step back to synthesize and summarize the findings related to the \experts{}' perceptions of AI's utility in mental healthcare and LLMs' performance handling ruptures. We identified six high-level reoccurring themes across the previous subsections wherein \experts{} had identified strengths and limitations: \textit{clinical task support, professional augmentation, access and availability, low-risk emotional support, safe space for self-disclosure,} and \textit{role boundaries and responsibilities} (see Table~\ref{tab:ai_role_comparison}).

\begin{table*}[tbh]
\footnotesize
\sffamily
\caption{\experts{}' broader perceptions of AI chatbots in mental health counseling mapped to rupture-related concerns and resolution implications.}
\label{tab:ai_role_comparison}
\setlength{\tabcolsep}{2pt}
\renewcommand{\arraystretch}{1.00}
\resizebox{\columnwidth}{!}{
\begin{tabular}{
>{\raggedright\arraybackslash}p{2.00cm}
>{\raggedright\arraybackslash}p{3.45cm}
>{\raggedright\arraybackslash}p{3.45cm}
>{\raggedright\arraybackslash}p{3.35cm}
>{\raggedright\arraybackslash}p{3.35cm}
}
\toprule
\textbf{Theme} 
& \textbf{Perceived Opportunities} 
& \textbf{Perceived Risks and Limitations} 
& \textbf{Rupture-Related Concerns} 
& \textbf{Resolution Implication} \\
\midrule

Clinical task support
& Assessment, screening, triage, psychoeducation, and structured therapeutic exercises
& Limited clinical accountability and limited relational adaptation
& AI responses may move too quickly into intervention or engage only superficially with the core rupture
& Slow the pacing, acknowledge the rupture first, and then offer structured support \\
\midrule
Professional augmentation
& Support for \experts{} through documentation, decision support, or preparation for care
& Risk of over-reliance on AI outputs without sufficient clinical oversight
& AI repair may appear adequate at baseline but still miss context-specific clinical nuance
& Treat AI-generated repair as a first response that requires professional review for timing and relational fit \\
\midrule
Access and availability
& Expanded access for underserved populations, therapy-hesitant users, and people seeking support before formal therapy begins
& Risk of deepening social isolation or replacing human connection with AI interaction
& AI may move the conversation forward while avoiding deeper exploration of the rupture
& Maintain continuity while inviting fuller reflection or connection to human support \\
\midrule
Low-risk emotional support
& Scalable and consistent support for routine or low-risk interactions
& Safety risks for vulnerable users when needs become more complex or acute
& AI may provide a reasonable first response but fail when distress becomes more complex
& Preserve clear boundaries and escalate or redirect when routine support is insufficient \\
\midrule
Safe space for self-disclosure
& A lower-barrier space for candid self-disclosure, especially for users hesitant to speak with others
& Data privacy, confidentiality concerns, and unknown long-term consequences
& User trust may weaken if the response feels mechanical, scripted, or based on unwarranted assumptions
& Use specific validation and avoid assuming emotions or needs that the user has not expressed \\
\midrule
Role boundaries and responsibilities
& Bounded support roles, such as navigator, coach, psychoeducational assistant, or reflection partner
& Unclear AI roles, professional displacement concerns, and inappropriate therapeutic authority
& AI may avoid escalation but limit therapeutic progress through overly safe or generic repair
& Clarify the system's role while still engaging the rupture meaningfully \\

\bottomrule
\end{tabular}}
\end{table*}

%% file: sections/6_Discussion.tex
\section{Discussion}

This study reveals a gap between what LLMs do well and what clinical practice requires. LLMs matched reference rupture labels at near ceiling agreement, while \experts{} rated the same models' \repair{} strategies as only moderately effective. 
We first outline why strong identification agreement should not be mistaken for genuine clinical insight (Section~\ref{sec:discussion_aggregate}), and what this gap means for designing rupture identification and \repair{} into future systems (Section~\ref{sec:discussion_design}).
We then discuss how such support should be bounded and reviewed (Section~\ref{sec:discussion_bounded}), and how this framing points toward broader Human-AI collaboration in mental health care (Section~\ref{sec:discussion_collaboration}).

\subsection{High Agreement on Rupture Identification Does Not Mean Clinical Understanding}
\label{sec:discussion_aggregate}

High agreement on rupture identification should not be interpreted as evidence of clinical understanding. As Sections~\ref{sec:findings_identification}, \ref{sec:findings_resolution}, and \ref{sec:findings_evaluation} show, LLMs often matched reference rupture labels but relied heavily on explicit linguistic cues rather than deeper interpretation of relational dynamics. This gap was clearer in resolution and evaluation tasks. \Experts{} rated the models' \repair{} strategies as only moderately effective and identified recurring problems, including premature problem solving, shallow engagement, and mechanical tone. Clinical interpretation therefore requires more than selecting the reference category.

Prior work has shown that strong performance on structured or aggregate metrics can overstate model capability when tasks require contextual judgment, domain expertise, and clinically meaningful interpretation~\cite{hicks2022evaluation, nori2025sequential, anand2026high}. Anand et al. similarly show that foundation models may perform well in aggregate on clinical cognitive assessment, while fine-grained evaluation reveals important gaps between model behavior and human judgment~\cite{anand2026high}. Our findings extend this concern to AI-mediated mental health conversations. Whereas LLMs tended to rely on explicit cues within individual turns, \experts{} integrated implicit, relational, and contextual signals across the broader conversation. They also treated ambiguity as central to clinical interpretation rather than as a problem to eliminate. This difference may reflect that LLMs are optimized to model textual patterns~\cite{bender2020climbing, sap2022neural}, while rupture \repair{} requires sensitivity to timing, interaction history, and relational context~\cite{song2025typing, moore2025expressing}.

Figure~\ref{fig:strategy_distribution} shows why aggregate summaries are insufficient. Averaged strategy distributions can make LLMs and \experts{} appear similar, but disaggregated views reveal that LLMs repeatedly relied on a narrow subset of strategies, while \experts{} used a broader and more scenario-sensitive range. The key question is therefore not whether a model selects plausible strategies on average, but whether it can adapt to the relational meaning of each interaction. This has implications for evaluating mental health AI systems. As Hullman argues, readers cannot judge whether an AI result is meaningful from aggregate metrics alone; they need access to tasks, examples, raw outputs, failure cases, baselines, prompts, interfaces, and design choices~\cite{hullman2026showmescience}. For rupture \repair{}, benchmarks should report not only scores or strategy distributions, but also representative rupture scenarios, model responses, expert rationales, and failure cases. Such materials are needed to assess whether a system responds to relational meaning or merely matches surface-level cues. More broadly, mental health AI benchmarks should evaluate how systems handle ambiguity, infer emotional meaning, and adapt responses based on interactional context. This aligns with HCI and healthcare AI research showing that clinical usefulness depends on context, workflow fit, and expert oversight, not isolated performance metrics alone~\cite{li2025vital, burgess2023healthcare, zajkac2024depends, zhang2024rethinking}.

\subsection{Designing Rupture Identification and \Repair{} into Future Systems}
\label{sec:discussion_design}
Future systems should treat rupture identification and \repair{} as fundamentally different design problems. Rupture identification lends itself to structured specification. Providing the 3RS marker definitions in the prompt allowed all three models to match reference labels across every high-level category and maintain near-ceiling agreement at the subtype level (Section~\ref{sec:findings_identification}). \Experts{} exhibited lower agreement and often interpreted subtle interpersonal tensions across the broader dialogue rather than anchoring on explicit lexical cues in individual turns, and they frequently treated ambiguous or overlapping cues as an inherent part of clinical interpretation rather than a problem to resolve immediately. 
Systems can leverage models' strength at applying explicit guidelines by supplying structured definitions to a standalone detection module that processes accumulated conversation history, since models tend to evaluate turns in isolation while clinicians track longitudinal dynamics. When cues are ambiguous or overlapping, however, the system should mirror \experts{}' interpretive stance by surfacing that uncertainty or asking clarifying questions rather than prematurely forcing a single label~\cite{sayres2026towards}. 

Rupture resolution, in contrast, cannot be reduced to a rubric. Clinicians draw on relational history and contextual judgment to decide how to respond to a particular person at a particular moment, and there is no universal mapping from rupture type to repair strategy. \Experts{} rated LLM-generated \repair{} strategies as only moderately effective, with recurring concerns about timing, depth, and contextual fit (Section~\ref{sec:findings_resolution}). LLMs moved prematurely toward problem-solving, missing the validation and relational attunement that \experts{} applied adaptively. A system that treats resolution as a completion task will appear competent on aggregate metrics while failing at precisely the moments that matter most.
This distinction suggests three design principles for future evaluation.

\begin{enumerate}
\item \textit{Prioritize validation over premature resolution.} Systems could delay problem-solving until the user's account has been explicitly elicited, heard, and acknowledged.
\item \textit{Preserve repair trajectory across dialogue states.} Systems could track previously attempted strategies across turns to reduce repetitive loops and support context-sensitive responses.
\item \textit{Consider human-in-the-loop handoff for unresolved strain.} Systems could transfer the dialogue to a human clinician when relational markers persist across turns despite repair attempts~\cite{haran2026checklist, moore2025expressing}.
\end{enumerate}

\noindent Prior research on clinician-informed interaction frameworks~\cite{thieme2023designing} and scenario-based simulation environments~\cite{adler2024beyond} offers practical methodologies for building and bounding these multi-turn resolution pathways prior to clinical deployment.

\subsection{Positioning LLMs as Bounded Support for Rupture Resolution}
\label{sec:discussion_bounded}

Given these limitations, LLMs should be positioned as bounded support rather than independent agents for rupture \repair{} and mental health care in general. As shown in Sections~\ref{sec:findings_evaluation} and~\ref{sec:outtake_survey}, \experts{} did not view AI as replacements for therapists~\cite{moore2025expressing} or as tools suited to relationally complex moments without expert oversight~\cite{jin2025applications}. They instead saw value in constrained forms of support~\cite{lawrence2024opportunities}. This boundary matters because systems that mimic emotional attunement without clinical judgment may appear to repair ruptures while leaving relational harm unaddressed. Recent lawsuits and safety discussions further show that emotionally responsive chatbots may worsen crises among vulnerable users when safeguards, escalation, and human review are insufficient~\cite{garcia2024character, ortutay2025openai, patel2024harmful, brittain2026character, godoy2025openai}. Together, these findings suggest moving away from therapist imitation and toward limited, reviewable, and context-bounded support.

LLMs were strongest when mapping explicit textual cues to predefined categories under controlled conditions, where inputs, outputs, and evaluation criteria were clearly specified. By contrast, tasks requiring timing, relational fit, and ongoing adaptation across turns showed the clearest gap between LLM outputs and expert judgment in Tables~\ref{tab:llm_repair_threeway} and~\ref{tab:ai_role_comparison}. This pattern suggests a division of labor in which LLMs support routine, preparatory, or structurally defined tasks, while experts remain responsible for relational interpretation and resolution. In rupture-specific contexts, systems could surface possible rupture markers, summarize relevant interaction histories, track recurring relational patterns, support preliminary assessment, or generate alternative response drafts for clinician review~\cite{yang2024talk2care, zhang2024rethinking, burgess2023healthcare, shi2025mapping, casu2024ai}. These uses remain bounded because the model supports preparation and reflection without deciding what the rupture means or how it should be resolved.

Designing for this boundary means making LLM support reviewable rather than decisive. Systems should present candidate rupture markers, surface uncertainty, and invite clinicians to interpret the interaction in context. This keeps \experts{} in the decision-making loop~\cite{buccinca2021trust, vasconcelos2023explanations} while allowing LLMs to support routine work that preserves clinicians' time and attention for patients~\cite{yang2024talk2care, thieme2023designing, adler2024beyond}. Future systems should make role boundaries explicit, communicate uncertainty, and provide mechanisms for escalation or human review. This shifts the design question from whether LLMs can imitate therapists to how they can support rupture-related workflows while preserving human responsibility for clinical judgment~\cite{adler2024beyond, englhardt2024classification}. In this role, LLMs can extend professional capacity without replacing the relational judgment at the center of mental health care~\cite{zou2026mind, zhao2026balancing, yang2024talk2care}.

\subsection{From Rupture Support to Collaborative Mental Health Workflows}
\label{sec:discussion_collaboration}

The bounded-support framing clarifies the limits of LLM use in rupture \repair{} and points to a broader design opportunity. Mental health AI should be developed not as a replacement for clinical or patient work, but as infrastructure for collaborative care. Rather than asking only whether LLMs can perform existing clinical tasks, future systems should ask how AI can help clinicians and patients prepare for, coordinate, and reflect on care.

This reframing follows from \experts{}' broader concerns about AI in mental health. As shown in Table~\ref{tab:ai_role_comparison}, perceived opportunities such as clinical task support, expanded access, and low-risk emotional support were closely tied to rupture-related risks, including premature intervention, shallow engagement, and scripted responses that miss underlying relational strain. Rupture is therefore not only a failure case for AI systems, but also a useful site for studying how AI-mediated care should be organized when relational quality becomes uncertain.

The more important design question is how AI can support the work around care rather than take over the work of care. Just as high agreement with reference labels does not establish clinical understanding, automating a narrow task does not necessarily improve care. A chatbot may classify a rupture, generate a plausible response, or draft documentation while still failing to support the relational and practical work that mental health care requires. Human-AI collaboration should therefore be evaluated by how well it improves preparation, continuity, reflection, and coordination across the care process.

One promising direction is clinician-facing workflow support. AI systems could reduce administrative burden and improve preparation by generating draft notes, summarizing clinically relevant themes, and organizing follow-up items for clinician review~\cite{yang2024talk2care, zhao2026balancing, zou2026mind}. Beyond reducing workload, such systems could make between-session patterns visible at the point of care, helping clinicians revisit concerns dispersed across notes, messages, and prior sessions ~\cite{chopra2025engagements, zou2026mind}. For rupture specifically, AI could support longitudinal review by surfacing recurring interaction patterns, flagging possible rupture markers across sessions, and organizing alternative response options for reflection. Scenario-based systems could also support training and supervision by helping trainees practice responses to rupture-like moments and compare strategies before entering higher-stakes settings~\cite{thieme2023designing, adler2024beyond}.

A second direction is patient-facing support for participation between sessions. Guided tools could help patients practice CBT skills such as reframing, grounding, emotion labeling, or behavioral activation while keeping clinicians informed about what patients practiced and where they struggled~\cite{fitzpatrick2017delivering}. AI could also support journaling and self-reflection by helping users organize experiences, notice recurring patterns, and prepare questions for future clinical conversations~\cite{chopra2025engagements, kim2024mindfuldiary, kim2024diarymate}. For users who face barriers to formal care, including cost, stigma, or geographic access, these systems may serve as lower-barrier entry points that connect users to professional support rather than substituting for it~\cite{lee2023exploring, inkster2018empathy}. In this role, AI would not provide therapy independently, but could help patients arrive at clinical encounters with clearer goals, examples, and questions.

Across these directions, the design challenge shifts from making AI appear therapeutically competent to making collaboration legible and useful. Systems should help users understand how outputs were produced, what evidence they draw on, and how they can be used, revised, or ignored in human care practices. In patient-facing contexts, this also means designing handoffs from reflection or skill practice to human support when users' needs exceed the system's intended role~\cite{haran2026checklist, moore2025expressing}. These handoffs are especially important given documented harms from systems that simulate relational understanding without clinical judgment~\cite{garcia2024character, ortutay2025openai, patel2024harmful, brittain2026character, godoy2025openai}.

Taken together, these findings point to a broader research agenda for mental health AI. Future work should move beyond evaluating whether LLMs can perform isolated clinical tasks and instead examine how they can support collaborative workflows that improve care quality, patient experience, and clinical practice. For HCI, this reframes mental health AI as a problem of collaborative workflow design, with rupture serving as a revealing test case for what relational competence in AI-mediated care requires.

%% file: sections/7_conclusion.tex
\section{Conclusion}
In this paper, we examined how LLMs identify and resolve conversational ruptures through a scenario-driven empirical study with 22 mental health \experts. Our findings showed that strong agreement with reference rupture labels did not translate into effective rupture resolution by LLMs. They often identified ruptures by mapping explicit linguistic cues to predefined categories, while \experts{} interpreted ruptures through relational and contextual cues. In resolution, \experts{} used adaptive strategies such as validation, open-ended exploration, psychoeducation, and collaboration, whereas LLM-generated resolution strategies were often more scripted, directive, and limited in timing, depth, and contextual sensitivity often stopping at just validation and agreement with the patient. 


These findings show that rupture handling in mental health AI cannot be reduced to agreement with reference labels or well-formed response generation.
Effective rupture resolution requires pacing, uncertainty awareness, relational fit, and adaptation over time. By grounding evaluation in psychotherapy theory, this work shows why clinically grounded perspectives are necessary for assessing LLMs in mental health interactions. Rather than positioning LLMs as replacements for therapists, our findings point to bounded human-AI collaboration that supports clinicians and patients while making system limits clear.
As these findings are based on constructed, text-based scenarios rather than naturally occurring conversations, further work is needed to confirm whether they extend to live interactions. 

%% file: sections/appendix.tex
\section*{Appendix}

This appendix provides additional details about the prompt used for LLMs. 
Appendix ~\ref{appendix:rupture_ident_prompt} presents the prompt used to identify rupture types in scenarios. 
Appendix ~\ref{appendix:rupture_resolution_generation_prompt} provides the prompt used to select resolution strategies and generate next chatbot responses.

\renewcommand{\thesubsection}{\Alph{subsection}}

\subsection{Prompt Used for Rupture Type Identification}
\label{appendix:rupture_ident_prompt}

\begin{lstlisting}
PROMPT_RUPTURE_TYPE_IDENTIFICATION = """
You are identifying the rupture type from the Chatbot--User conversation.
First, select one or more of the 6 high-level rupture categories that apply.
Then, for each selected category, select one subtype within that category that best fits the conversation.
Finally, explain in one sentence per selected subtype why it represents that rupture.
If multiple ruptures are present, list them clearly. Do not include any other explanation or comments.

High-level rupture categories and subtypes

1. Disengagement
- Minimal response

2. Avoidance
- Abstract communication
- Avoidant storytelling or shifting topic
- Denial

3. Incongruence
- Deferential or appeasing
- Content--affect split

4. Control
- Patient rejects therapist intervention
- Patient defends self against therapist
- Efforts to control or pressure the therapist

5. Self-Directed Negativity
- Self-criticism or hopelessness

6. Dissatisfaction
- Complaints or concerns about the therapist
- Complaints or concerns about the activities of therapy
- Complaints or concerns about the parameters of therapy
- Complaints or concerns about progress in therapy

Definitions

Minimal response is when ``the patient withdraws from the therapist by going silent or by giving minimal responses to questions or statements that are intended to initiate or continue discussion,'' and these minimal responses ``function to shut down the therapist's attempts to engage the patient in the work of therapy.''

Abstract communication is when ``the patient avoids the work of therapy by using vague or abstract language,'' and this abstract language ``functions to keep the therapist at a distance from the patient's true feelings, concerns, or issues.''

Avoidant storytelling and shifting topic is when ``the patient tells stories and/or shifts the topic in a manner that functions to avoid the work of therapy,'' sometimes shutting the therapist out ``as if the patient were not even aware that the therapist is there.''

Denial is when ``the patient withdraws from the therapist and/or the work of therapy by denying a feeling state that is manifestly evident, or denying the importance of interpersonal relationships or events that seem important and relevant to the work of therapy,'' and this denial ``functions to shut down or move away from the current topic or activity, thereby hindering the work of therapy.''

Deferential and appeasing behavior is when ``the patient withdraws from the therapist and/or the work of therapy by being overly compliant and submitting to the therapist in a deferential manner,'' in a way that ``functions to avoid conflict with the therapist'' and ``makes it harder for the therapist to know how the patient really feels or what the patient really thinks.''

Content/affect split is when ``the patient withdraws from the therapist and/or the work of therapy by exhibiting affect that does not match the content of his/her narrative.''

Self-criticism and hopelessness is when ``the patient withdraws from the therapist and the work of therapy by becoming absorbed in a depressive process of self-criticism and/or hopelessness that seems to shut out the therapist and to close off any possibility that the therapist or the treatment can help the patient.''

Patient rejects therapist intervention is when ``the patient rejects or dismisses the therapist's intervention,'' attacking or shutting down ``something that the therapist is trying to bring to the table.''

Patient defends self against therapist is when ``the patient defends his/her thoughts, feelings, or behavior against what he/she perceives to be the therapist's criticism or judgment,'' making ``a case to support, validate, and defend his/her behavior, beliefs, feelings, decisions, etc.''

Efforts to control or pressure the therapist is when ``the patient attempts to control the therapist and/or the session, or the patient puts pressure on the therapist to fix the patient's problems quickly,'' including ``trying to push or provoke the therapist.''

Complaints or concerns about the activities of therapy is when ``the patient expresses dissatisfaction, discomfort, or disagreement with specific tasks of therapy such as homework assignments or in-session tasks.''

Complaints or concerns about the parameters of therapy is when ``the patient expresses concerns or complaints about the parameters of treatment, such as the therapy schedule\ldots\ session length, number and frequency of sessions, or the research contract.''

Complaints or concerns about progress in therapy is when ``the patient expresses complaints, concerns, or doubts about the progress that can be made or has been made in therapy.''

Complaints or concerns about the therapist is when ``the patient expresses negative feelings about the therapist,'' including feeling ``angry, impatient, distrustful, manipulated, hurt, judged, controlled, rejected,'' or feeling ``that the therapist has failed to support, encourage, or respect him/her.''

Chatbot--User conversation

C is the Chatbot and U is the User.
\{C: \ldots\ U: \ldots\}
"""
\end{lstlisting}

\subsection{Prompt Used for Resolution Strategy Generation}
\label{appendix:rupture_resolution_generation_prompt}

\begin{lstlisting}
PROMPT_RESOLUTION_STRATEGY_GENERATION = """
You are identifying the best resolution strategy from the Chatbot-User conversation. Please select the strategy that would be the most effective response for the AI assistant to use next. After selecting a strategy, briefly explain in one sentence why you chose it, and provide the chatbot's ideal next response in one sentence.

# 10 Resolution Strategy Options

Therapist clarifies a misunderstanding: "Therapist responds to a rupture by attempting to clarify a misunderstanding. Generally, the resolution effort stops here; the therapist does not go on to explore the underlying significance of the misunderstanding."

Therapist changes tasks or goals: "The therapist changes the tasks or goals of therapy in response to a rupture. The therapist may change the task/goal in order to address the concerns of a patient who is complaining about the tasks and goals of therapy, or the therapist may attempt to engage a withdrawn patient by changing the task or goal."

Therapist illustrates tasks or provides a rationale for treatment: "The therapist responds to a rupture by illustrating, explaining, or providing a rationale for a therapy task or goal. The therapist may share his/her reasons for pursuing a particular task or goal in an effort to engage the patient or alleviate concerns."

Therapist invites the patient to discuss thoughts or feelings with respect to the therapist or some aspect of therapy: "The therapist responds to a rupture by inviting the patient to express negative or vulnerable thoughts or feelings about the therapist and/or the tasks or goals of therapy."

Therapist acknowledges his/her contribution to a rupture: "The therapist acknowledges his/her contribution to a rupture."

Therapist discloses his/her internal experience of the patient-therapist interaction: "The therapist discloses his/her internal experience of the patient-therapist interaction."

Therapist links the rupture to larger interpersonal patterns between the patient and the therapist: "The therapist links the rupture to larger interpersonal patterns between the patient and the therapist."

Therapist links the rupture to larger interpersonal patterns in the patient's other relationships: "The therapist links the rupture to larger interpersonal patterns in the patient's other relationships."

Therapist validates the patient's defensive posture: "The therapist validates the patient's defensive posture."

Therapist responds to a rupture by redirecting or refocusing the patient: "The therapist responds to a rupture by redirecting or refocusing the patient."

# Output format

Resolution Strategy: <One of the 10 strategies exactly as written>

Reason: <Why this strategy is the most effective in one or two sentences>

Next Chatbot Response: <The ideal next message>

# Chatbot-User conversation

C is the Chatbot and U is the User.
{C: ... U:...}
"""
\end{lstlisting}